\documentclass[conference]{IEEEtran}
\IEEEoverridecommandlockouts
\usepackage{cite}
\usepackage{amsmath,amssymb,amsfonts}
\usepackage{algorithmic}
\usepackage{graphicx}
\usepackage{textcomp}
\usepackage{xcolor}
\usepackage{multirow}
\usepackage{url}
\usepackage{hyperref}
\usepackage{threeparttable}
\usepackage{booktabs, siunitx, threeparttable, caption, subcaption}
\usepackage{mathtools}
\usepackage{makecell}
\usepackage{comment}
\usepackage{float}
\def\BibTeX{{\rm B\kern-.05em{\sc i\kern-.025em b}\kern-.08em
    T\kern-.1667em\lower.7ex\hbox{E}\kern-.125emX}}
\begin{document}

\title{\LARGE \bf i-Tac: Inverse Design of 3D-Printed Tactile Elastomers with \\Scalable and Tunable Optical and Mechanical Properties




\author{Wen Fan, Dandan Zhang*}

\thanks{W. Fan and D. Zhang are with the Department of Bioengineering, Imperial College London, UK (emails: \{d.zhang17, w.fan24\}@imperial.ac.uk). * Corresponding author: D. Zhang (email: d.zhang17@imperial.ac.uk).}
}

\maketitle

\begin{abstract}
Elastomers are central to vision-based tactile sensors (VBTSs), where they transduce external contact into observable deformation. Different VBTS architectures, however, require distinct optical and mechanical properties, particularly transparency and hardness. Conventional elastomer design relies on a forward, trial-and-error optimisation process from material preparation to property evaluation, which is inefficient and offers limited property scalability and target tunability.

In this work, we present \textit{i-Tac}, an inverse design pipeline for tailoring 3D-printed tactile elastomers with target optical and mechanical properties. Inspired by the composite structure of the human dermis, i-Tac exploits multi-material PolyJet additive manufacturing with three complementary resins. A mixture design methodology is employed to characterise the printed elastomers and establish response surface models (ReSMs) that map material compositions to functional properties, thereby defining a scalable property space. Based on user-defined targets, a desirability-function-based multi-objective optimisation is then performed to identify feasible composition regions and derive an optimal operating window for fabrication. This enables elastomers with desired properties to be manufactured in a single iteration, thereby achieving efficient target tunability. That is, the ability to systematically realise a specified material formulation from a given property target.

Experimental results validate the proposed i-Tac framework in terms of both property scalability and inverse design performance, showing that i-Tac can effectively tailor elastomer transparency and hardness while reducing the iterative burden of conventional forward design. By fabricating physical sensor samples from both commercial and custom designs, the proposed framework further demonstrates the potential of inverse-designed, monolithically manufactured elastomers for customisable VBTS fabrication.
\end{abstract}


\begin{figure}[!htbp]
	\centering
	\includegraphics[width = 1\hsize]{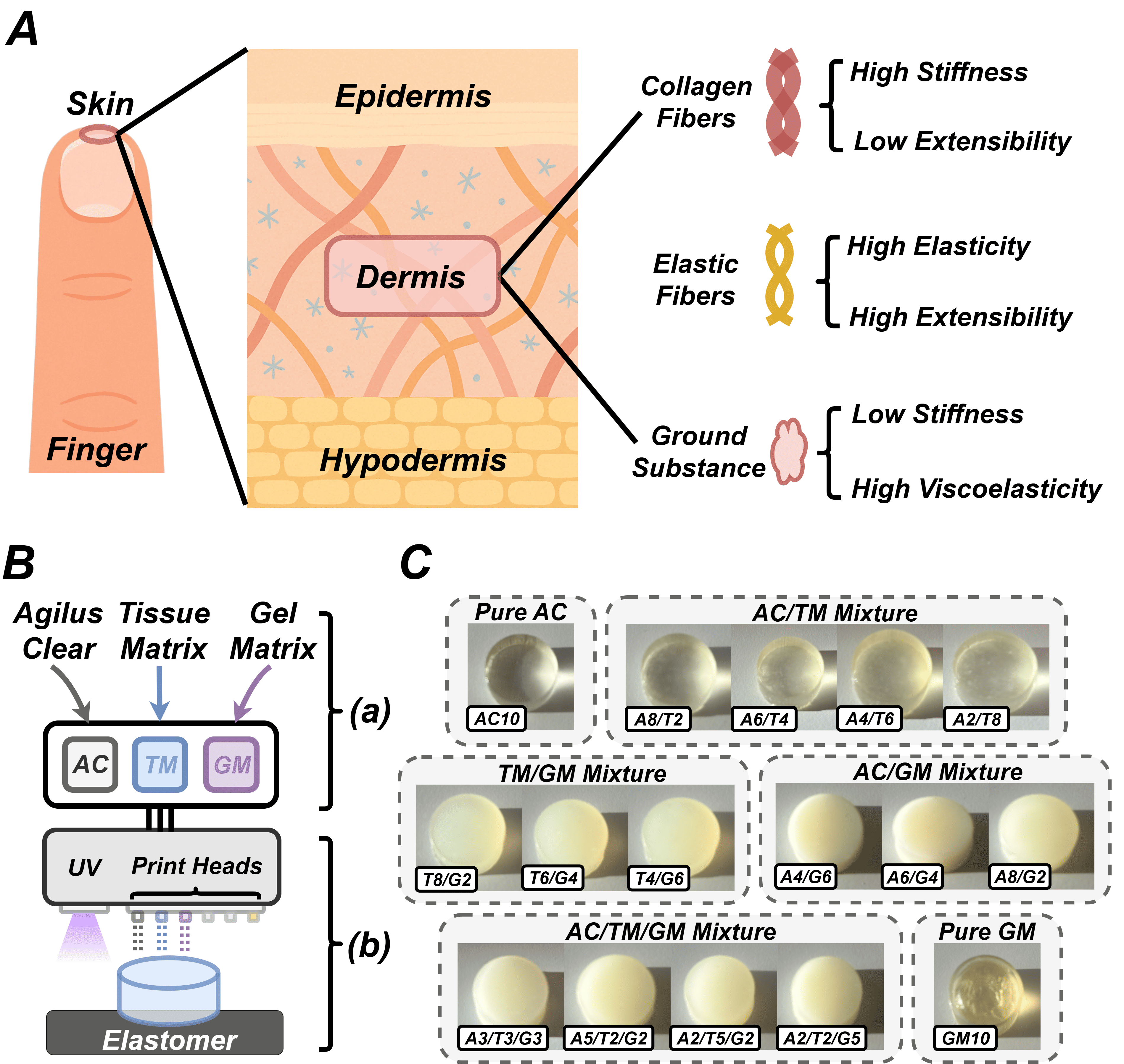}
	\caption{\small A: Physiological anatomy of human skin, in which the dermis consists of three distinct tissue components. B: Inverse design framework for achieving target tunability in elastomer properties through (a) AC/TM/GM mixture design and (b) monolithic manufacturing. C: Scalable property space of AC/TM/GM mixture elastomers, covering a broad range from clear to opaque and from soft to rigid.} 
	\label{bio principle}
\end{figure}

\begin{figure*}[!htbp]
	\centering
	\includegraphics[width = 1\hsize]{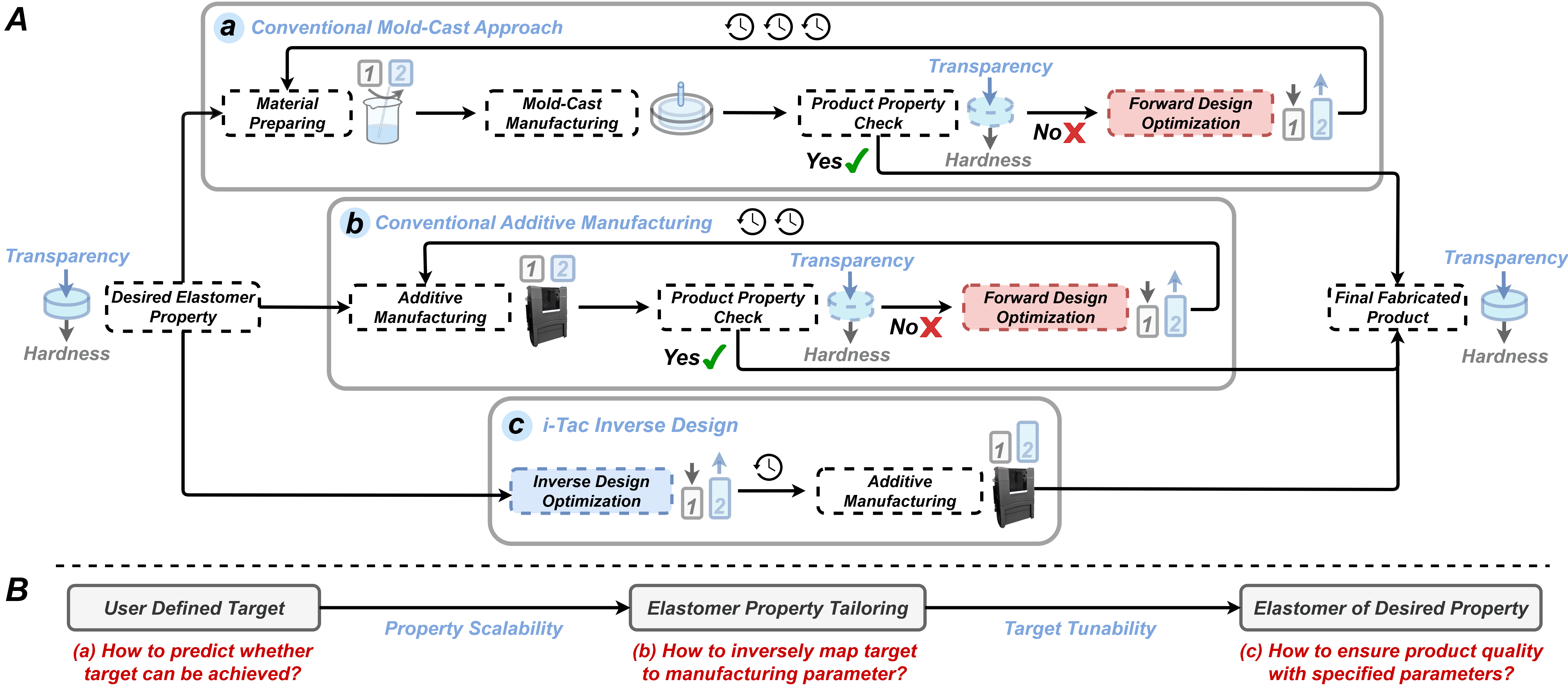}
	\caption{\small A: Comparison between forward design and inverse design paradigm in elastomer property tailoring. (a)/(b) Forward design requires prototyping in advance of parameter modification, though additive manufacturing can accelerate fabrication. (c) Inverse design of i-Tac helps additive manufacturing in rapid elastomer tailoring without iteration. B: Challenges to be addressed in elastomer property tailoring.} 
	\label{framework compare}
\end{figure*}

\section{Introduction}

Elastomers provide compliance for flexible sensing and inherent safety in physical interaction, and are therefore widely used in vision-based tactile sensors (VBTSs)~\cite{zhang2022hardware}. In VBTSs, external contact stimuli first induce deformation of the elastomeric interface~\cite{fan2024vitactip}. The sensing mechanism then converts this deformation into measurable visual cues, such as pixel intensity variations and marker displacements, which are captured by the embedded vision system~\cite{fan2025crystaltac}. 
Consequently, the optical and mechanical properties of the elastomer play a decisive role in this transduction process, with transparency and hardness being two of the most critical properties for tactile sensing.

Across different VBTS designs, the required elastomer properties of transparency and hardness span a broad spectrum, ranging from optically clear to translucent~\cite{lin20239dtact,cong2025taceva,zhang2023tirgel}, and from soft to stiff~\cite{yuan2017gelsight,zhang2025design,gomes2020geltip}. This diversity gives rise to two fundamental requirements for VBTS elastomer design. First, the elastomer system should exhibit sufficient \textbf{property scalability}, such that its achievable optical and mechanical property range is broad enough to cover diverse target requirements. Second, once a target property has been confirmed to lie within this feasible range, the elastomer system should also support \textbf{target tunability}, namely, the ability to systematically identify and realise the material formulation required to achieve that target. These two requirements correspond to target feasibility and target realisability, respectively.

In general, silicone elastomers are widely used as compliant interaction media in VBTSs, offering a broad range of optical and mechanical properties through variations in mixture ratio, curing conditions, and additional chemical components. 
However, the conventional mold-casting approach faces important limitations. First, it is highly sensitive to manufacturing variables, making it difficult to ensure reliable and reproducible elastomer properties. Second, it relies on a forward design paradigm in which the manufacturing parameters can only be tuned towards the desired target after the actual elastomer properties have been evaluated.


As illustrated in Fig.~\ref{framework compare} (A.a), this trial-and-error optimisation process requires repeated prototyping, making target-oriented tuning inefficient. Although physics-based or semi-empirical models can be employed to pre-calibrate the properties of mold-cast VBTS elastomers~\cite{di2024using}, such approaches require blending silicone phases with similar curing conditions and therefore offer limited scalability for multi-objective property customisation. Ultimately, mold-casting approach affects VBTSs not only by prolonging the development cycle, but also by introducing performance variability into the finished products. 


Recently, \textbf{additive manufacturing} has been introduced into VBTS fabrication~\cite{fan2024vitactip} as a promising alternative for enabling faster production, with 3D-printed elastomers being investigated alongside mold-cast samples for tactile sensing~\cite{solayman2024mechanical}. Furthermore, studies on monolithic manufacturing~\cite{fan2024design, fan2024magictac, fan2025crystaltac} have explored the fabrication of entire sensor contact modules, including the elastomer, through multi-material additive manufacturing. However, these studies have primarily emphasised flexibility in shaping elastomer geometries, such as structural dimensions or nested configurations within the sensor, while research into elastomer property scalability remains limited. Moreover, the target tunability of 3D-printed elastomer properties, such as transparency and hardness, also relies on repeated cycles of production and modification within a forward design pipeline, as shown in Fig.~\ref{framework compare} (A.b).

To overcome these limitations, we seek to establish a new approach inspired by \textbf{biological principles}. In human tactile sensing\cite{willemet2022biomechanics}, the skin serves as the primary transduction interface, and its physiological structure provides valuable insights for elastomer design. In Fig.~\ref{bio principle} (A), skin exhibits a multilayered architecture: (1) epidermis forms a thin protective layer of keratinocytes; (2) hypodermis, composed mainly of adipose tissue, connects to deeper muscles and bones; (3) between them lies the dermis, which governs the skin’s mechanical behaviour. The dermis is itself a composite of collagen fibres (70\%), elastic fibres (4\%), and ground substance (20\%), interspersed with mechanoreceptors responsible for tactile sensing. Collagen fibres provide high stiffness but limited extensibility, elastic fibres are both highly extensible and elastic, and the fluidic ground substance is soft and viscoelastic. 
The interplay of these distinct components gives rise to the superior biomechanical properties of human skin. This architecture not only ensures robustness and adaptability under diverse mechanical conditions, but also demonstrates how integrating materials with complementary properties can yield functional performance that exceeds that of any single constituent.

\begin{figure*}[!htbp]
	\centering
	\includegraphics[width = 1\hsize]{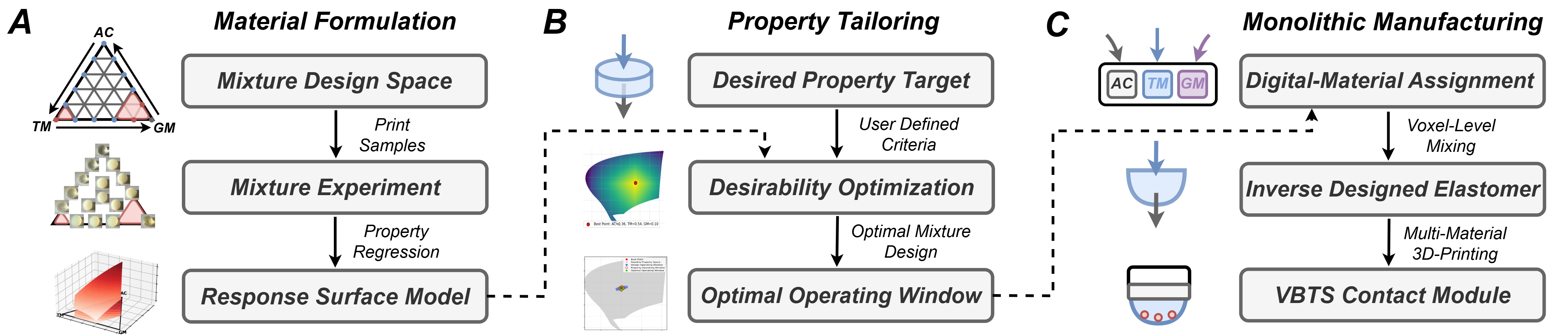}
	\caption{\small i-Tac inverse design pipeline for elastomer property tailoring in VBTSs. A: Material formulation, construction of response surface models to map AC/TM/GM mixture designs to elastomer properties and establish a scalable property space. B: Property tailoring, identification of feasible operating windows for 3D printing from desired property targets using a desirability function, enabling target tunability. C: Monolithic manufacturing, realisation of the inverse-designed elastomer in a sensor via multi-material printing.} 
	\label{inverse design pipeline}
\end{figure*}

Motivated by this biological mixture principle, we propose to achieve property scalability in VBTS elastomers by combining 3D-print materials with complementary properties. As shown in Fig.~\ref{bio principle} (B), three PolyJet printing materials are introduced: Agilus30 Clear (AC)\footnote{\url{https://www.stratasys.com/en/materials/materials-catalog/polyjet-materials/agilus30/}}, TissueMatrix (TM)\footnote{\url{https://www.stratasys.com/en/materials/materials-catalog/polyjet-materials/tissuematrix/}}, and GelMatrix (GM)\footnote{\url{https://www.stratasys.com/en/materials/materials-catalog/polyjet-materials/gelmatrix/}}. Specifically, (1) AC is a rubber-like material with a Shore hardness of 30A and good transparency; (2) TM is considerably softer, with tissue-like properties and moderate transparency; and (3) GM exhibits gel-like characteristics with high fluidity. Following the principle observed in the dermis of human skin, elastomers fabricated through \textbf{AC/TM/GM mixture design} can span a wide spectrum of optical and mechanical properties, thereby providing a scalable property space for diverse VBTS requirements. As illustrated in Fig.~\ref{bio principle} (C), monolithic manufacturing~\cite{fan2025crystaltac} further maximises this potential through precise digital-material assignment rather than manual material preparation, enabling accurate voxel-level mixing to realise the above mixture-design concept in a manner analogous to the composite structure of the dermis.

Merely introducing an AC/TM/GM mixture design is insufficient to achieve property tailoring in VBTS elastomers. As analysed in Fig.~\ref{framework compare} (B), users would still face the intractable task of navigating a forward design process between multiple target properties and their corresponding material compositions. To address this unresolved challenge, an \textbf{inverse design approach} is required to enable efficient prediction from desired elastomer properties to optimal mixture configurations, thereby providing the target tunability needed to realise tailored elastomers in a single batch (Fig.~\ref{framework compare} (A.c)). The main challenge of such prediction, however, lies in the relatively high cost of elastomer property measurement, which constrains the collection of large-scale datasets for data-driven approaches. In this regard, statistical approaches, including \textbf{mixture experiments}~\cite{cornell2011retrospective}, \textbf{response surface models (ReSMs)}~\cite{khuri2010response}, and \textbf{desirability functions}~\cite{costa2011desirability}, provide suitable solutions, as they jointly offer a data-efficient and interpretable framework for inverse design under limited experimental data.

The proposed inverse design pipeline is illustrated in Fig.~\ref{inverse design pipeline}. First, a mixture design space is constructed for the AC/TM/GM experiment, and the optical and mechanical properties of the sampled mixtures are modelled using two ReSMs (Fig.~\ref{inverse design pipeline} (A)). Next, given the desired property targets with the ReSMs defined above, a desirability function is employed to perform multi-objective optimisation. Finally, the resulting optimal mixture design is translated into an operational window (Fig.~\ref{inverse design pipeline} (B)), which can be realised through monolithic manufacturing and used as an inverse-designed elastomer in VBTSs  (Fig.~\ref{inverse design pipeline} (C)). 

The main contributions of this work are as follows:
\begin{itemize}
    \item We propose an \textbf{AC/TM/GM mixture design} using three complementary printing materials via a multi-material additive manufacturing framework. Inspired by the tissue-mixing principle of the human finger dermis, this design significantly expands the achievable property range of PolyJet 3D-printed elastomers in terms of both optical transparency and mechanical hardness, thereby enhancing their property scalability.
    \item We propose an \textbf{inverse design pipeline} that directly maps target properties to AC/TM/GM mixture configurations of 3D-printed elastomers. By avoiding the conventional forward design process, this approach provides an interpretable and efficient route for achieving target tunability in elastomer property design.
    \item We employ \textbf{monolithic manufacturing} to fabricate real samples with minimal property deviation. The experimental results validate the effectiveness of both the proposed mixture design and the inverse design pipeline. This strategy helps narrow the quality gap between 3D-printed sensors and conventionally fabricated VBTS products based on mold-cast silicone and manual assembly.

\end{itemize}

\section{Related Work}

\subsection{Manufacturing Methods for VBTS Elastomers}

Existing fabrication approaches for VBTS elastomers fall into two principal categories: mold casting and additive manufacturing~\cite{zhang2022hardware}. The majority of reported VBTSs employ the mold-casting approach, including both commercially available products and lab-based prototypes~\cite{yuan2017gelsight, gomes2020geltip, zhang2022deltact, lin20239dtact, zhang2023tirgel, fan2024vitactip}, as well as those produced by industry~\cite{lambeta2020digit}. Despite its prevalence, mold casting is constrained by its reliance on molds, labor-intensive processing, and variability in product quality~\cite{fan2024design}. To overcome these limitations, several studies have explored additive manufacturing techniques for VBTS elastomers, such as direct ink writing (DIW)~\cite{solayman2024mechanical} and PolyJet printing (PP)~\cite{fan2024magictac}. 

Compared with DIW, which relies on a limited number of printheads, PP employs arrays of printheads with fine nozzles, thereby enabling higher-resolution multi-material fabrication~\cite{fan2025crystaltac}. On this basis, PP enables precise blending of multiple printing materials through digital-material assignment. Such fine-scale mixing enhances homogeneity and mitigates the fluctuations in product quality that are often observed in mold-cast elastomers, where inaccuracies typically stem from manual material preparation. However, PP is restricted to proprietary PolyJet printing materials rather than widely used commercial silicones, thereby limiting its scalability for elastomer property tailoring. Recently, monolithic manufacturing~\cite{fan2024magictac, fan2025crystaltac} has demonstrated the potential of PP for achieving softer elastomers, but it still relies on multi-layer grid structures and remains limited in optical transparency. Building upon this line of research, our work advances monolithic manufacturing towards homogeneous elastomers through bio-inspired multi-material mixture design, thereby enhancing the property scalability of VBTS elastomers across both optical and mechanical dimensions.

\begin{figure*}[!htbp]
	\centering
	\includegraphics[width = 1\hsize]{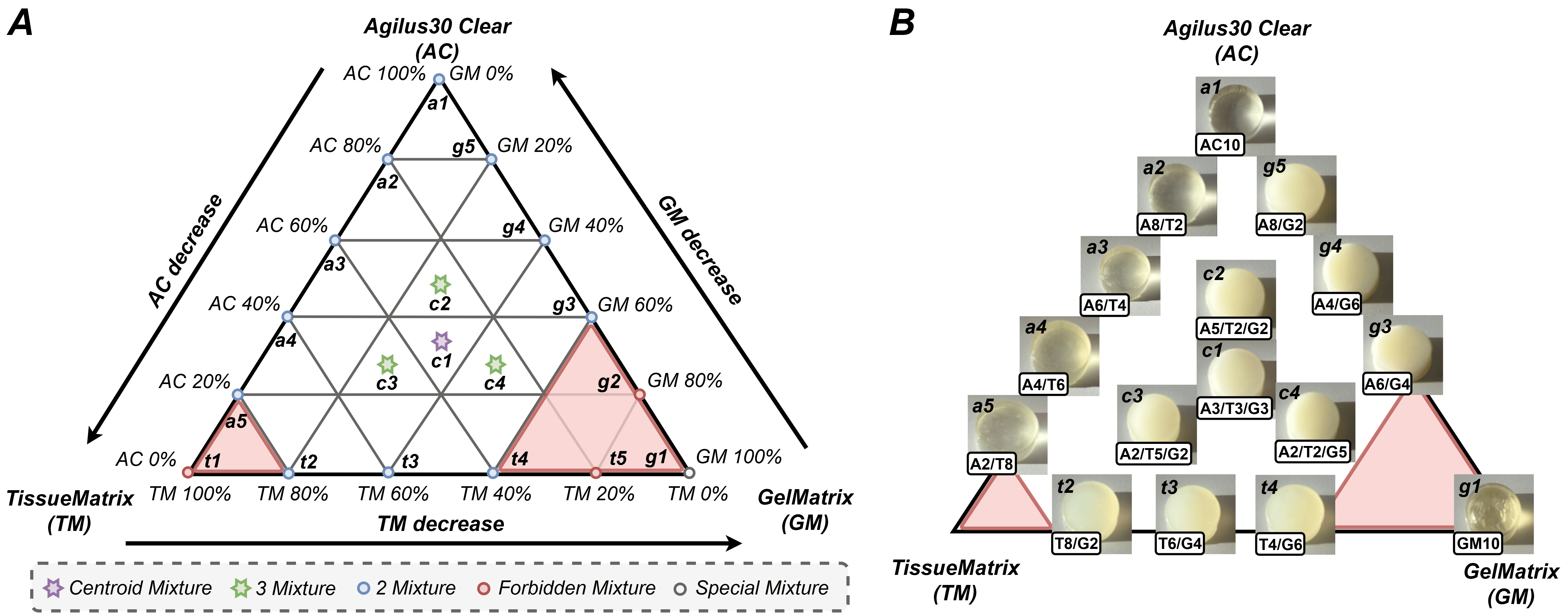}
	\caption{\small A: Mixture design space of AC/TM/GM component, with centroid mixture (c1), 3-mixture (c2-c4), 2-mixture (a1-a5,t2-t4,g3-g5), forbidden mixture (t1,t5,g2), and special mixture (g1). B: Fifteen samples (a1-a5,t2-t4,g1,g3-g5,c1-c4) are fabricated for mixture experiment.} 
	\label{mds}
\end{figure*}

\subsection{Inverse Design Methods for Desired Material Property}

In materials science, inverse design for achieving target material properties can be broadly categorised into four paradigms~\cite{han2025ai}: (1) experiment-driven; (2) theory-driven; (3) computation-driven; and (4) artificial intelligence (AI)-driven. Within additive manufacturing, machine learning has emerged as a prominent approach under the AI-driven paradigm, owing to its ability to deliver rapid and accurate predictions of printed material properties~\cite{ng2024progress}. Under this paradigm, recent studies have explored data-driven inverse design methods for tailoring the mechanical behaviour of 3D-printed metamaterials. Ha et al.~\cite{ha2023rapid} proposed a rapid inverse design framework that reproduces high-fidelity stress-strain responses using generative models. Chai et al.~\cite{chai2024tailoring} developed a machine-learning pipeline trained on finite-element data to programme snapping metamaterials with pronounced nonlinearities. 

However, data-driven machine-learning approaches generally require large datasets and are often less interpretable than explicit models~\cite{allen2022machine}. This limitation makes them less suitable for VBTS elastomer design, where data acquisition is expensive and labor-intensive. As an alternative, the statistical principles of mixture experiments~\cite{mclean1966extreme, cornell2011retrospective} enable the optimisation of material properties with relatively few samples, while also providing physical insight through explicit mixture design. For instance, Gao et al.~\cite{gao2016use} applied mixture experiments to investigate how the proportional compositions of hemicellulose, cellulose, and lignin affect product distributions during hydrothermal treatment. Similarly, in additive manufacturing, Zhang et al.~\cite{zhang2023hybrid} employed mixture experiments to optimise functional ink formulations for aerosol jet printing, using response surface methodology and desirability functions to balance conductivity and printability. Such an approach is well suited to VBTS elastomer design, as it supports data-efficient modelling of composition-property relationships, thereby providing a practical foundation for target tunability.

\begin{table}[htbp]
\centering
\small
\setlength{\tabcolsep}{5pt}
\renewcommand{\arraystretch}{1.12}
\caption{Summary of terms used in this paper}
\label{tab:symbols}
\begin{tabular}{@{} l >{\raggedright\arraybackslash}p{0.62\columnwidth} @{}}
\toprule
\textbf{Term} & \textbf{Definition} \\
\midrule
AC & Agilus30 Clear \\
TM & TissueMatrix \\
GM & GelMatrix \\
$x_1$ & AC component (\%) \\
$x_2$ & TM component (\%) \\
$x_3$ & GM component (\%) \\
$(x_1, x_2, x_3)$ & Mixture Design Space (MDS) \\
$Y(x)$ & Response Surface Model (ReSM) \\
$Y_1$ & ReSM of transparency (\%) \\
$Y_2$ & ReSM of hardness (Shore 00) \\
$Y^*_1,Y^*_2$ & Optimized ReSM after ANOVA \\
$x(Y_1,Y_2)$ & Feasible Property Space (FPS) \\
FPS (AC) & $x_1(Y_1,Y_2)$ \\
FPS (TM) & $x_2(Y_1,Y_2)$ \\
FPS (GM) & $x_3(Y_1,Y_2)$ \\
NTB & Nominal-The-Better desirability criteria \\
LTB & Larger-The-Better desirability criteria \\
STB & Smaller-The-Better desirability criteria \\
$T_{Y1},T_{Y2}$ & Desired target of transparency/hardness \\
$w_1,w_2$ & Weight for transparency/hardness \\
$d(Y)$ & Desirability function to reach $T_{Y}$ \\
$D(d(Y)...)$ & Overall desirability function \\
$(\hat{x_1},\hat{x_2},\hat{x_3})$ & Optimal mixture design \\
$(\hat{Y_1},\hat{Y_2})$ &  ReSM prediction of $(\hat{x_1},\hat{x_2},\hat{x_3})$ \\
$W_{\Delta x}$ & Design operating window ($\pm\Delta x\%$) \\
$W_{\Delta Y}$ & Property operating window ($-\Delta Y\%$)\\
$W_{\text{optimal}}$ & Optimal operating window $\{(\hat{x}_1,\hat{x}_2,\hat{x}_3)_0,\ldots,(\hat{x}_1,\hat{x}_2,\hat{x}_3)_i\}$ \\
$R_{Y1},R_{Y2}$ & Real transparency/hardness of elastomer \\
Error1 & Error between $T_Y$ and $\hat{Y}$ \\
Error2 & Error between $\hat{Y}$ and measured $R_{Y}$ \\
\bottomrule
\end{tabular}
\end{table}

\section{Methodology}

For clarity, the principal symbols, abbreviations, and notations used in the following methodology and experiments are summarised in Table \ref{tab:symbols}.

\subsection{Mixture Design Space}


Following the mixture experiment framework \cite{cornell2011retrospective}, the problem is formulated as an n-component mixture (n = 3):
\begin{equation}
\begin{aligned}
    0\le L_i \le x_i \le U_i \le 1, i=1,2,...,n\\
\end{aligned}
\end{equation}
where \(L_i\) and \(U_i\) denote the lower and upper bounds of each material component \(x_i\) due to printer settings, where $x_1 \in [0\%, 100\%], x_2 \in [0\%, 80\%], x_3 \in [0\%, 60\%]$, also satisfying \(\sum_{i=1}^{n} x_i = 1\), with each component percentage restricted to integer values \(x_i \in \{0\%, 1\%, 2\%, \dots, 100\%\}\). In general, the composition space of \(n\) components corresponds to an \((n-1)\)-dimensional simplex, thereby \textbf{mixture design space (MDS)} of AC/TM/GM reduces to an equilateral triangle $(x_1, x_2, x_3)$, represented as a \textbf{ternary diagram} in Fig.~\ref{mds} (A).




\subsection{Mixture Experiment}

\subsubsection{Selected Samples}

From MDS, a total of 19 mixtures can be obtained through uniform sampling. Among these, c1 corresponds to the centroid mixture, c2-c4 represent three-component mixtures, while the remainder are binary mixtures. Owing to the lower and upper bounds of each material, t1, t5, and g2 are excluded as forbidden mixtures, whereas g1 is isolated for GM optical testing due to its distinctive fluidity. Consequently, 15 samples (a1-a5, t2-t4, g1, g3-g5, c1, c2-c4) are selected for following mixture experiment, with full details provided in Table~\ref{15 sample point} and illustrated in Fig.~\ref{mds} (B).

\begin{table}[!htbp]
    \centering
    \caption{\small Selected Samples for Mixture Experiment}
    \label{15 sample point}
    \begin{tabular}{|c|c|c|c|c|c|}
    \hline
        \textbf{Type} & \textbf{Sample} & \textbf{Mixture Design} & \textbf{$x_1$} & \textbf{$x_2$} & \textbf{$x_3$} \\ \hline
        \multirow{11}{*}{\textbf{2-mixture}} 
        & \textbf{a1} & AC10 & 100\% & 0 & 0 \\ \cline{2-6}
        & \textbf{a2} & AC8/TM2 & 80\% & 20\% & 0 \\ \cline{2-6}
        & \textbf{a3} & AC6/TM4 & 60\% & 40\% & 0 \\ \cline{2-6}
        & \textbf{a4} & AC4/TM6 & 40\% & 60\% & 0 \\ \cline{2-6}
        & \textbf{a5} & AC2/TM8 & 20\% & 80\% & 0  \\ \cline{2-6}
        & \textbf{t2} & TM8/GM2 & 0 & 80\% & 20\% \\ \cline{2-6}
        & \textbf{t3} & TM6/GM4 & 0 & 60\% & 40\%\\ \cline{2-6}
        & \textbf{t4} & TM4/GM6 & 0 & 40\% & 60\%\\ \cline{2-6}
        & \textbf{g3} & AC4/GM6 & 40\% & 0 & 60\% \\ \cline{2-6}
        & \textbf{g4} & AC6/GM4 & 60\% & 0 & 40\% \\ \cline{2-6}
        & \textbf{g5} & AC8/GM2 & 80\% & 0 & 20\% \\ \hline
        \multirow{4}{*}{\textbf{3-mixture}} 
        & \textbf{c1}  & A3/T3/G3 & 33\% & 33\% & 34\%\\ \cline{2-6}
        & \textbf{c2} & A5/T2/G2 & 50\% & 25\% & 25\%\\ \cline{2-6}
        & \textbf{c3} & A2/T5/G2 & 25\% & 50\% & 25\% \\ \cline{2-6}
        & \textbf{c4} & A2/T2/G5 & 25\% & 25\% & 50\%\\ \hline
    \end{tabular}
\end{table}

\subsubsection{Optical Property Experiment}

\begin{figure}[!htbp]
	\centering
	\includegraphics[width = 1\hsize]{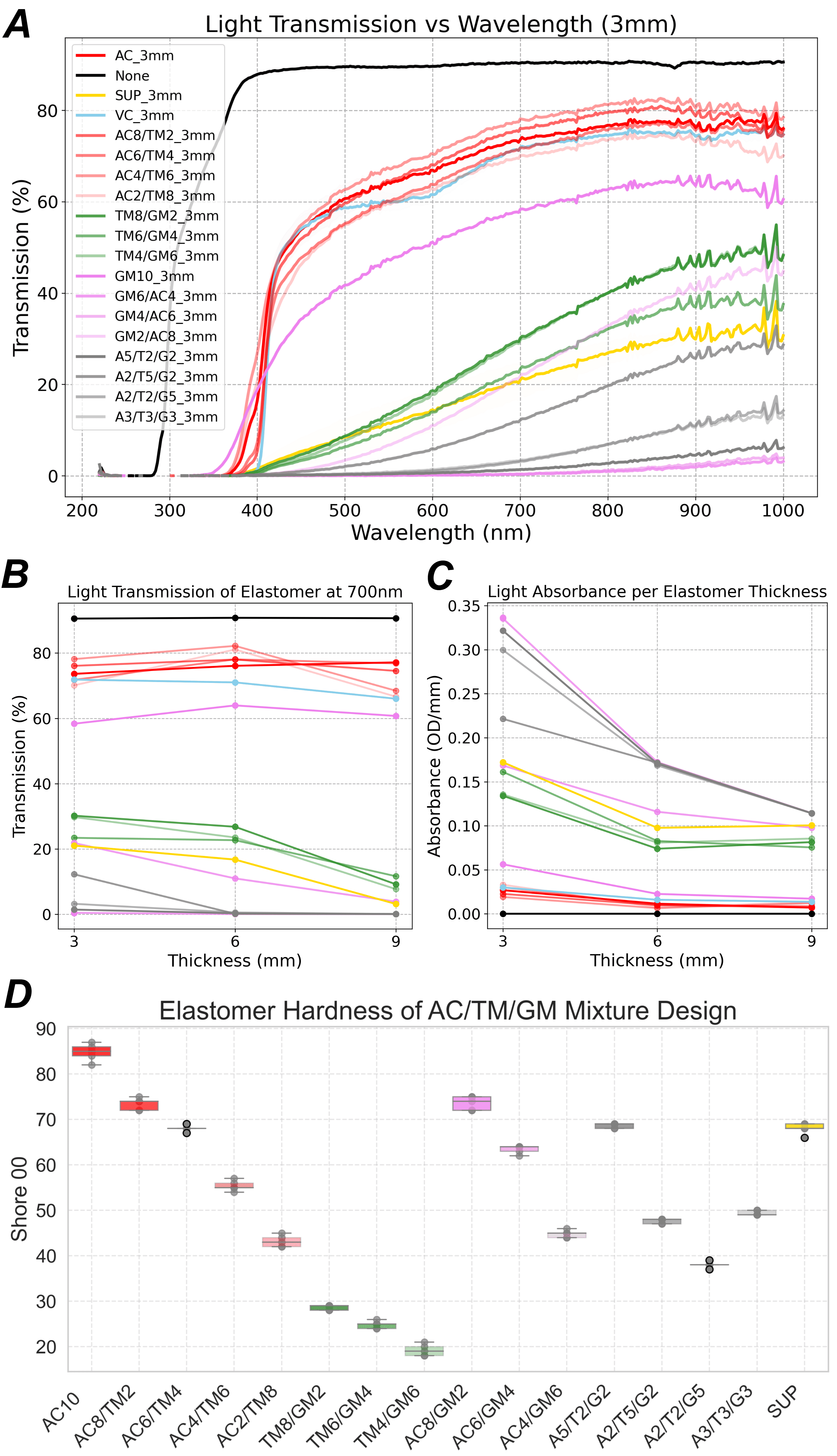}
	\caption{\small A: Light transmission per wavelength of mixture design samples with 3mm thickness. B: Light transmission at 700nm with different sample thickness. C: Opacity density per unit thickness. D: Mechanical hardness of mixture design samples.} 
   \label{optical_mechanical_property_result}
\end{figure}

 \begin{figure*}[!htbp]
	\centering
	\includegraphics[width = 1\hsize]{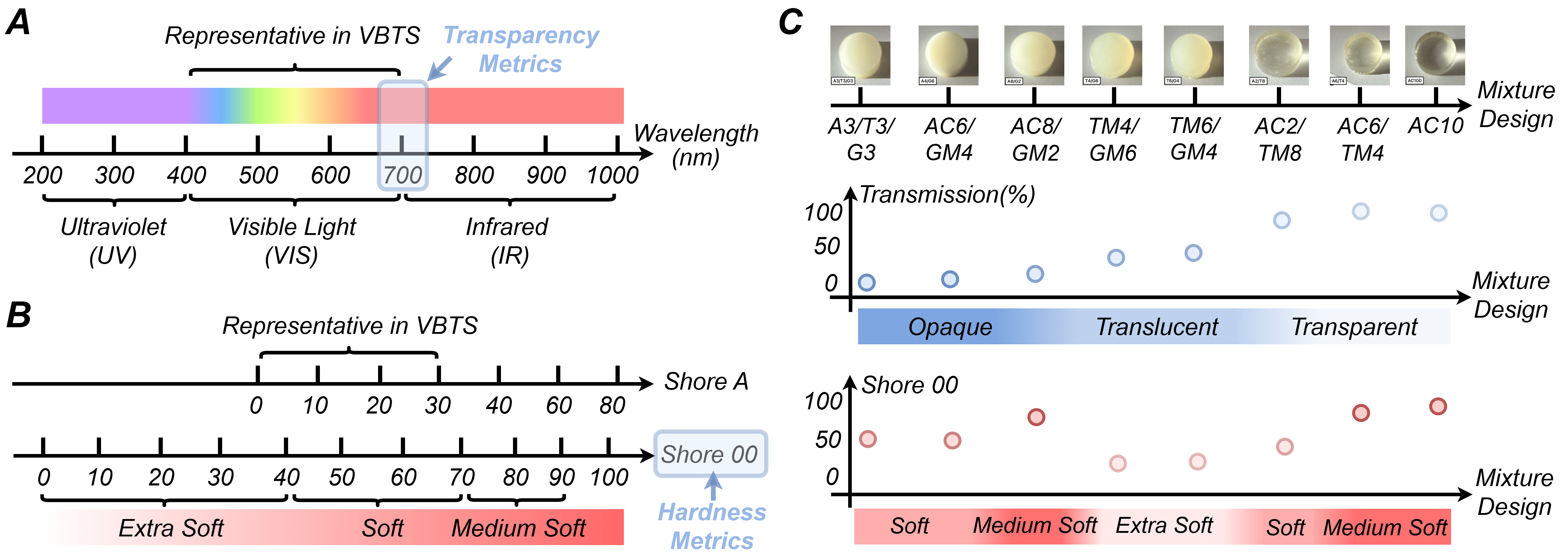}
	\caption{\small A: Transparency metrics in optical mixture experiment. B: Hardness metrics in mechanical mixture experiment. C: Experiment results in terms of optical transparency and Shore 00 hardness demonstrate wide property scalability of mixture designs.} 
	\label{wavelength_shore 00}
\end{figure*}

To quantitatively analyse the optical transparency of the 3D-printed elastomer, light transmission \(T\) as a function of wavelength \(\lambda\) is introduced, defined as the ratio of input to output light intensities:  

\begin{equation}
    T(\lambda) = \frac{I_{\text{out}}(\lambda)}{I_{\text{in}}(\lambda)}
\end{equation}

Based on the 15 mixture-design samples, a total of 90 cylindrical specimens were fabricated, each with elastomer thicknesses \(d_{\text{thick}} = 3, 6, 9\,\text{mm}\) and a diameter of 10 mm, where a 1 mm surface layer of VeroClear was added upon to ensure structural rigidity. All specimens were subsequently placed in a microplate and tested using a spectrometer with a scanning range of \(\lambda \in [200,1000]\,\text{nm}\), covering the ultraviolet (200-400 nm), visible (400-700 nm), and infrared (700-1000 nm) regions.  In addition, multi-layer grid (SUP) \cite{fan2024magictac,fan2025crystaltac}, VeroClear (VC), and pure air (None) were included as baselines. As an illustrative case, the measured transmission \(T(\lambda)\) at \(d_{\text{thick}} = 3\,\text{mm}\) is shown in Fig.~\ref{optical_mechanical_property_result} (A), demonstrating that the light transmittance of the elastomers generally increases with wavelength, while exhibiting marked variations across different mixture designs. To achieve efficient property scalability and target tunability, \(T(\lambda)\) is too complex to be a concise indicator, therefore, \(T_{\text{mix}}\) is defined  for each mixture design as its light transmission at a particular wavelength \(\lambda\) for a given elastomer thickness \(d_{\text{thick}}\). 


As most VBTSs operate with visible-light sources, the optical properties within the 400–700 nm range are the most relevant. In these applications, the design prioritises overall transmittance over detailed spectral shaping.
Thereby, \(\lambda = 700\,\text{nm}\) is selected, as it provides the highest transmission of each mixture design within the visible range (Fig.~\ref{optical_mechanical_property_result} (B)). On this basis, the relationship between \(T_{\text{mix}}\) and \(d_{\text{thick}}\) can be examined using the Beer-Lambert law \cite{swinehart1962beer}, which relates absorbance \(A\), path length \(d\), extinction coefficient \(\epsilon\), and concentration \(c\) of an absorbing substance:

\begin{equation}
    A= -\log_{10}(T) = \epsilon \, c \, d
\end{equation}

By setting the path length \(d\) equal to \(d_{\text{thick}}\) and introducing a bias transmission \(T_{\text{bias}}\) measured from pure-air baseline, the absorbance of mixture design and opacity density per unit thickness can be mathematically expressed as:

\begin{equation}
    A_{\text{mix}} = -\log_{10}\!\big(T_{\text{mix}} + (1 - T_{\text{bias}})\big) = c_{\text{opacity}} \, d_{\text{thick}}
\end{equation}

\begin{equation}
    c_{\text{opacity}} = \frac{\log_{10}\!\big(T_{\text{mix}} + (1 - T_{\text{bias}})\big)}{d_{\text{thick}}}
\end{equation}

Finally, \(c_{\text{opacity}}\) characterises the effective ‘opacity density per unit thickness’ of the mixture design samples. As shown in Fig.~\ref{optical_mechanical_property_result} (C), \(c_{\text{opacity}}\) at \(d_{\text{thick}} = 3\,\text{mm}\) exhibits a more uniform distribution compared with the clustered convergence observed at 6 mm and 9 mm, indicating improved differentiation of the mixture-design characteristics.



\subsubsection{Mechanical Property Experiment}

Similarly, Shore 00 hardness is employed to characterise the mechanical properties of the mixture designs. In contrast to the optical specimens, larger specimens with a diameter of 30 mm and a thickness of 12 mm were fabricated for the mechanical experiments. For each sample, five hardness measurements were uniformly collected from the surface using a Shore 00 durometer, yielding a total of 75 data points across 15 mixture designs. The results are presented in Fig.~\ref{optical_mechanical_property_result} (D), which reflect a linear relationship between elastomer hardness and the mixture compositions.




\subsubsection{Mixture Experiment Result}



\begin{table}[!htbp]
    \centering
    \caption{\small Mixture Experiment Results of Transparency/Hardness}
    \begin{tabular}{|c|c|c|}
    \hline
         \textbf{Mixture Design} & \textbf{Unbiased Transparency} & \textbf{Hardness (Shore 00)} \\ \hline
         AC10 & 83.06\% & 84.8 \\ \hline
         AC8/TM2 & 85.25\% & 73.4 \\ \hline
         AC6/TM4 & 81.23\%  & 68.0 \\ \hline
         AC4/TM6 & 87.59\%  & 55.4 \\ \hline
         AC2/TM8 & 79.56\%  & 43.2 \\ \hline
         TM8/GM2 & 39.61\%  & 28.6 \\ \hline
         TM6/GM4 & 32.83\%  & 24.8 \\ \hline
         TM4/GM6 & 39.19\%  & 19.2 \\ \hline
         AC4/GM6 & 9.78\%  & 44.8 \\ \hline
         AC6/GM4 & 9.85\%  & 63.4 \\ \hline
         AC8/GM2 & 31.24\%  & 73.6 \\ \hline
         A3/T3/G3 & 42.73\%  & 49.4 \\ \hline
         A5/T2/G2 & 10.84\%  & 68.6 \\ \hline
         A2/T5/G2 & 21.66\%  & 47.6 \\ \hline
         A2/T2/G5 & 12.62\%  & 38.0 \\ \hline
         GM10 & 67.79\%  & Gel-like \\ \hline
         VC & 81.3\%  & Rigid \\ \hline
         SUP & 30.46\% & 68.0 \\ \hline
         Pure Air & 100.0\%  & - \\ \hline
         
    \end{tabular}
    \label{table sample data}
\end{table}

\begin{figure*}[!htbp]
	\centering
	\includegraphics[width = 1\hsize]{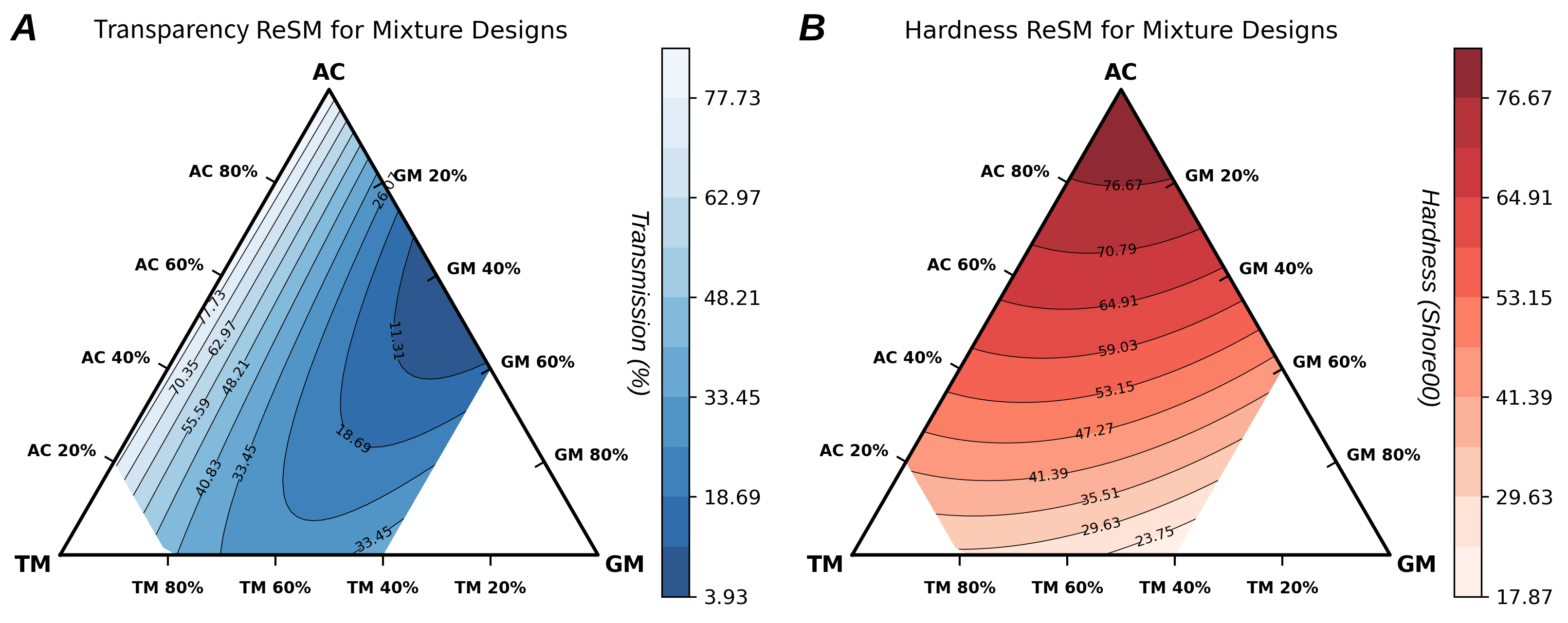}
	\caption{\small A: Transparency ReSM for mixture designs. B: Hardness ReSM for mixture designs} 
	\label{ReSM_2d}
\end{figure*}

As shown in Fig.~\ref{wavelength_shore 00} (A), transmission \(T_\text{{mix}} = T(700nm)\) is recorded as the optical transparency data of mixture design sample (\(d_{\text{thick}} = 3\,\text{mm}\)), representing its most clarity within visible light range. From Fig. \ref{optical_mechanical_property_result} (A), the mixtures exhibit near-linear, monotonic spectra with peak transmittance around 700 nm. Therefore, the single-wavelength metric preserves the relative ordering of candidate materials for property scalability and target tunability. As shown in Table~\ref{table sample data}, it can be further corrected as unbiased transparency by \(T_{\text{bias}}\) of pure-air measurement. From Fig.~\ref{wavelength_shore 00} (B), 
the Shore 00 scale extends to softer materials than Shore A, since its measurement range is better suited to the hardness values of AC/TM/GM mixtures.


Based on these two criteria, example results of mixture experiments are illustrated in Fig.~\ref{wavelength_shore 00} (C) and detailed in Table~\ref{table sample data}. With respect to optical properties, all AC/TM mixtures exhibit high transparency of 70-80\%, whereas TM/GM mixtures show significantly reduced transmission, ranging from 30\% down to 15\%. Other designs, such as AC/GM and three-component mixtures, rapidly become opaque, with transmission falling below 10\% or approaching zero. Regarding mechanical hardness, values decrease linearly from Shore 00-85 to 00-40 as AC fraction is reduced from 100\% to 20\%, a trend also observed for AC/GM mixtures. The addition of TM or GM further softens the samples compared with AC-based designs, with TM/GM mixtures reaching as low as Shore 00-30 or even 00-20. Collectively, these results demonstrate that AC/TM/GM mixtures expand both optical and mechanical property scalability of 3D-printed elastomers, also highlighting their strong potential for inverse design in target tunability.

\begin{table*}[!ht]
  \centering
  \caption{ANOVA Optimisation of Response Surface Model}
  \label{tab:anova}
  \begin{threeparttable}
  \small
  \begin{tabular}{l *{10}{c}}

    \toprule
    & \multicolumn{2}{c}{$Y_1$}
    & \multicolumn{2}{c}{$Y^*_1$}
    & \multicolumn{2}{c}{$Y_2$}
    & \multicolumn{2}{c}{$Y^*_2(-x_3)$}
    & \multicolumn{2}{c}{$Y^*_2(+x_3)$} \\
    \cmidrule(lr){2-3}\cmidrule(lr){4-5}\cmidrule(lr){6-7}\cmidrule(lr){8-9}\cmidrule(lr){10-11}
    \multicolumn{1}{c}{} & F & P & F & P & F & P & F & P & F & P \\
    \midrule
    $x_1$   & 73.992 & $<0.0001$ & 150.149 & $<0.0001$ & 1018.021 & $<0.0001$ & 1199.306 & $<0.0001$ & 1288.709 & $<0.0001$ \\
    $x_2$   & 15.207 & 0.0045    & 73.575  & $<0.0001$ & 31.445   & $<0.0001$ & 44.940   & $<0.0001$ & 43.334   & 0.0001 \\
    $x_3$   & 6.452  & 0.0347    & 14.487  & 0.0034    & 0.603    & \textbf{0.4597} & --- & --- & 2.085 & 0.183 \\
    $x_{12}$  & 0.194  & \textbf{0.6712} & --- & --- & 2.029 & 0.1921 & 5.651 & 0.0388 & 6.983 & 0.0268 \\
    $x_{13}$  & 13.059 & 0.0068    & 34.690  & 0.0002    & 6.357    & 0.0357 & 37.341 & 0.0001 & 16.697 & 0.0027 \\
    $x_{23}$  & 3.836  & 0.0859    & 15.179  & 0.0030    & 2.347    & 0.1641 & 9.994  & 0.0101 & 8.531  & 0.0170 \\
    $x_{123}$ & 0.294  & \textbf{0.6021} & --- & --- & 0.461 & \textbf{0.5159} & --- & --- & --- & --- \\
    \midrule
    RMSE $\downarrow$
         & \multicolumn{2}{c}{10.605}
         & \multicolumn{2}{c}{\textbf{9.664}}
         & \multicolumn{2}{c}{2.886}
         & \multicolumn{2}{c}{2.946}
         & \multicolumn{2}{c}{\textbf{2.798}} \\
    $R^2$ $\uparrow$
         & \multicolumn{2}{c}{0.979}
         & \multicolumn{2}{c}{0.978}
         & \multicolumn{2}{c}{0.999}
         & \multicolumn{2}{c}{0.998}
         & \multicolumn{2}{c}{0.998} \\
    Adj.\,$R^2$ $\uparrow$
         & \multicolumn{2}{c}{0.960}
         & \multicolumn{2}{c}{0.967}
         & \multicolumn{2}{c}{0.997}
         & \multicolumn{2}{c}{0.997}
         & \multicolumn{2}{c}{0.997} \\
    AIC $\downarrow$
         & \multicolumn{2}{c}{117.980}
         & \multicolumn{2}{c}{114.540}
         & \multicolumn{2}{c}{78.935}
         & \multicolumn{2}{c}{78.902}
         & \multicolumn{2}{c}{77.777} \\
    BIC $\downarrow$
         & \multicolumn{2}{c}{122.936}
         & \multicolumn{2}{c}{118.080}
         & \multicolumn{2}{c}{83.892}
         & \multicolumn{2}{c}{82.442}
         & \multicolumn{2}{c}{82.026} \\
    \bottomrule
  \end{tabular}

  \vspace{0.4em}
  \begin{tablenotes}
    \footnotesize\centering
    \item \textbf{Note:} ‘$Y^*_2(-x_3)$’ represents $Y^*_2$ without $x_3$; ‘$Y^*_2(+x_3)$’ indicates $Y^*_2$ with reintroduced $x_3$.
  \end{tablenotes}
  \end{threeparttable}
\end{table*}

\subsection{Response Surface Model (ReSM)}

\subsubsection{Quadratic Regression Formulation}

Based on the experimental data summarised in Table~\ref{table sample data}, a response surface model (ReSM) is employed to perform regression fitting for transparency and hardness, thereby scaling the limited sample measurements from mixture design space to the full property prediction surface. In particular, the cubic Scheffé model \cite{cornell1991fitting} is adopted to characterise the relationships between transparency/hardness \((Y_1/Y_2)\) and the volume fractions of the mixture-design components \((x_1, x_2, x_3)\):

\begin{equation}
    Y_m = \sum_{i=1}^{n} \beta_i^m x_i \;+\; \sum_{1 \leq i < j}^{n} \beta_{ij}^m x_i x_j \;+\; \sum_{1 \leq i < j < k}^{n} \beta_{ijk}^m x_i x_j x_k
\end{equation}

where \(n=3\) corresponds to AC/TM/GM components, \(m \in [1,2]\) represents the transparency/hardness properties, and \(\beta\) denotes the regression coefficients. The terms \(x_i\), \(x_i x_j\), and \(x_i x_j x_k\) represent the main effects (AC, TM, GM), the two-way interactions (AC/TM, AC/GM, TM/GM), and the three-way interaction (AC/TM/GM), respectively. The regression coefficients \(\beta\) are estimated using the ordinary least squares (OLS) method:



\begin{equation}
\hat{\beta} 
= \arg\min_{\beta} \, \| y - X\beta \|^2
= (X^\top X)^{-1} X^\top y
\end{equation}

Based on the fitted parameters \(\hat{\boldsymbol{\beta}}\), the regression models for transparency \((Y_1)\) and hardness \((Y_2)\) are obtained as follows:

\begin{equation}
\begin{aligned}
  Y_1 &= 82.5\,x_1 + 71.9\,x_2 + 110.6\,x_3 + 27.1\,x_1x_2 \\
      &\quad - 362.1\,x_1x_3 - 233.2\,x_2x_3 - 167.6\,x_1x_2x_3
\end{aligned}
\end{equation}

\begin{equation}
\begin{aligned}
  Y_2 &= 83.3\,x_1 + 28.1\,x_2 - 9.2\,x_3 + 23.9\,x_1x_2 \\
      &\quad + 68.7\,x_1x_3 + 49.6\,x_2x_3 + 57.1\,x_1x_2x_3
\end{aligned}
\end{equation}

\begin{figure*}[!htbp]
	\centering
	\includegraphics[width = 1\hsize]{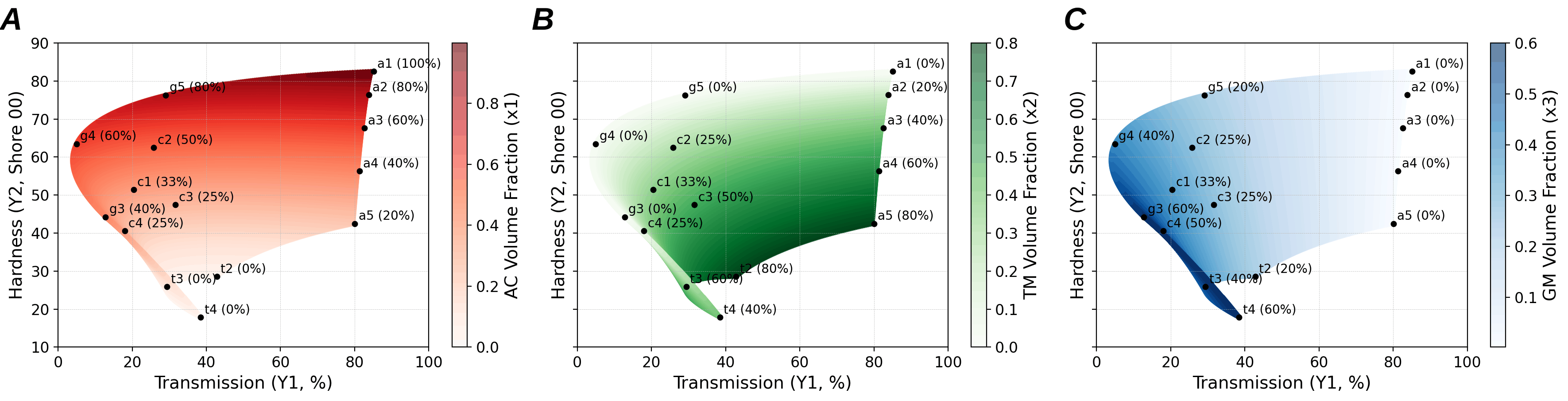}
	\caption{\small Feasible property space. A: FPS(AC), $x_1(Y_1,Y_2)$, effects more in higher hardness area. B: FPS(TM), $x_2(Y_1,Y_2)$, effects more in lower hardness area. C: FPS(GM), $x_3(Y_1,Y_2)$, effects more in lower clarity and ultra-soft area. } 
	\label{feasible space}
\end{figure*}

\subsubsection{ANOVA for Model Optimization}

Analysis of variance (ANOVA) is applied to refine the initial regression models \(Y_1\) and \(Y_2\) by removing independent factors with negligible influence on the dependent variables, as summarised in Table~\ref{tab:anova}. For \(Y_1\), the \(p\)-values of \(x_{12}\) and \(x_{123}\) are significantly higher than those of the other terms, indicating that the AC/TM and three-way interactions have limited impact on optical transparency and can thus be omitted. Similarly, in the case of \(Y_2\), the terms \(x_{3}\) and \(x_{123}\) yield the largest \(p\)-values, suggesting that GM and the three-way interaction contribute less to mechanical hardness. Two optimised regression models, ${Y^*_1}$ and ${Y^*_2}$, are therefore proposed, with their performance evaluated in Table~\ref{tab:anova}. For ${Y^*_1}$, model simplification by excluding \(x_{12}\) and \(x_{123}\) results in lower RMSE (greater predictive accuracy), higher \(R^2\) or adjusted \(R^2\) (indicating improved variable selection), and lower AIC/BIC (more concise formulation). In contrast, ${Y^*_2}$ achieves greater simplicity (lower AIC/BIC) but suffers from reduced prediction accuracy (higher RMSE) when omitting \(x_{3}\) and \(x_{123}\). Since GM plays a critical role, \(x_{3}\) is reintroduced, leading to improved regression precision compared with the initial \(Y_2\).  The final optimised models are thus denoted as \({Y_1}^*\) and \({Y_2}^*\):


\begin{equation}
\begin{aligned}   
   {Y^*_1} &= 85.1\,x_1 + 78.8\,x_2 + 124.2\,x_3 \\
      &\quad  - 399.1\,x_1x_3 - 281.6\,x_2x_3 
\end{aligned}
\end{equation}


\begin{equation}
\begin{aligned}   
  {Y^*_2} &= 82.5\,x_1 + 26.2\,x_2 - 13.7\,x_3  \\
      &\quad + 31.6\,x_1x_2 + 81.0\,x_1x_3 + 65.1\,x_2x_3
\end{aligned}
\end{equation}

\begin{figure*}[!htbp]
	\centering
	\includegraphics[width = 1\hsize]{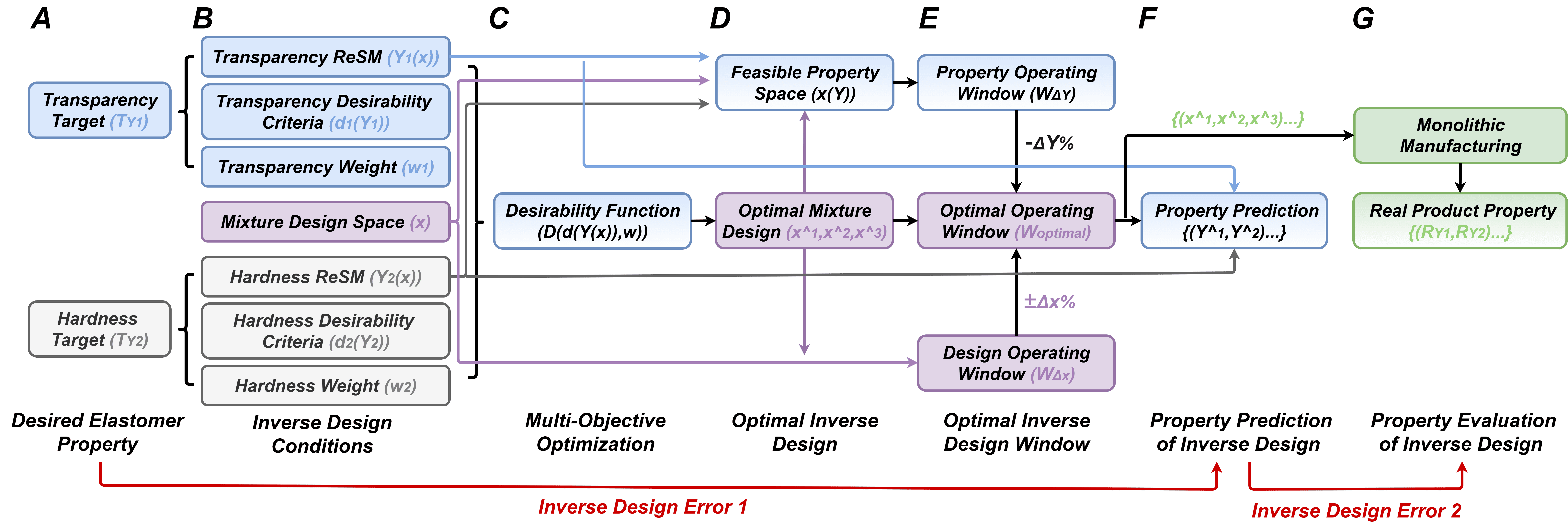}
	\caption{\small Inverse design methodology for tailoring elastomer property. A: Desired elastomer property. B: Required material formulations and user preference. C: Multi-objective optimization between different property targets. D: Optimal inverse design. E: Optimal inverse design window for elastomer fabrication. F: Property prediction of inverse designed elastomer. G: Real evaluated property of elastomer product.} 
	\label{plot:inverse design method}
\end{figure*}

\subsubsection{ReSM Analysis}

Based on ANOVA optimization results, two corresponding ReSMs, $Y_1 = Y^*_1(x_1,x_2,x_3)$ and $Y_2 = Y^*_2(x_1,x_2,x_3)$, are respectively constructed to enable an intuitive \textbf{material-property formulation} from mixture designs $x_1/x_2/x_3$ to elastomer properties $Y_1/Y_2$, with their 2D contour maps illustrated in Fig.~\ref{ReSM_2d}. For the transparency ReSM (Fig.~\ref{ReSM_2d} (A)), its response surface exhibits a non-monotonic variation: higher proportions of AC ($x_1$) and TM ($x_2$) generally lead to greater transparency ($Y_1$) of the printed elastomer. Moreover, when the GM ($x_3$) content remains below 10\%, $Y_1$ consistently exceeds 55\%. In contrast, the hardness ReSM (Fig.~\ref{ReSM_2d} (B)) presents a more homogeneous distribution. The vertex occurs at pure AC ($x_1=100\%$), and hardness gradually decreases along the axis extending toward TM/GM ($x_2$/$x_3$). The softest region of $Y_2$ lies along the TM/GM edge, indicating an inverse relationship between hardness $Y_2$ and the AC fraction ($x_1$).

\begin{table*}[!ht]
    \centering
    \caption{Desirability Criteria Guideline of Inverse Design for Tailoring Elastomer Property}
    \begin{tabular}{|c|c|c|c|c|c|c|}
    \hline
        \textbf{Guidance Id} & $Y_1$ \textbf{Criteria} & $Y_2$ \textbf{Criteria} & \textbf{Tailoring Elastomer Property} & \textbf{Inverse Design Application} \\ \hline
        Id-1  & NTB & NTB & Specific elastomer transparency and hardness & Targeted elastomer customization \\ \hline
        Id-2 & NTB & LTB & Hardest elastomer based on a specific transparency & Hardness optimization \\ \hline
        Id-3 & NTB & STB & Softest elastomer based on a specific transparency & Hardness optimization \\ \hline
        Id-4 & LTB & NTB & Clearest elastomer based on a specific hardness & Transparency optimization \\ \hline
        Id-5 & LTB & LTB & Clearest and hardest elastomer & Ultra-rigid clear elastomer \\ \hline
        Id-6 & LTB & STB & Clearest and softest elastomer & Ultra-soft clear elastomer \\ \hline
        Id-7 & STB & NTB & Most opaque elastomer based on a specific hardness & Transparency optimization \\ \hline
        Id-8 & STB & LTB & Most opaque and hardest elastomer & Ultra-rigid opaque elastomer \\ \hline
        Id-9 & STB & STB & Most opaque and softest elastomer & Ultra-soft opaque elastomer \\ \hline
    \end{tabular}
    \label{desirability criteria guidance}
\end{table*}

\subsubsection{Feasible Property Space}

To lay the foundation for subsequent inverse design tasks, three \textbf{feasible property spaces (FPSs)} are investigated, $x_1(Y_1,Y_2)$, $x_2(Y_1,Y_2)$, and $x_3(Y_1,Y_2)$, which reversely maps from $Y_1/Y_2$ to $x_1/x_2/x_3$, as illustrated in Fig.~\ref{feasible space}. It characterizes the scalable property distribution of mixture designs, whose outline represents the inverse design boundary on target tunability. Furthermore, the UV-cured nature and partially overlapping property ranges of AC, TM, and GM promote smooth property variation with mixing ratio, supported by the near-linear hardness trends observed at the feasible space boundaries (Fig. \ref{optical_mechanical_property_result}(D)). In addition, the convex-combination formulation of the mixture design physically bounds the achievable properties within the envelope defined by the pure constituents. This constraint reduces the risk of unexpected discontinuities.

A larger FPS area indicates broader elastomer property scalability achievable through mixture design, thereby providing greater inverse design generalisability. As shown in Fig.~\ref{feasible space}, the overall FPS exhibits an irregular property envelope with a parallelogram-shaped region on the right, indicating that the range of $Y_1$ between 40-80\%, and $Y_2$ between Shore 00-40 and 00-80 can be largely attained through appropriate combinations of $x_1/x_2/x_3$. In contrast, the left region of the FPS forms a triangular cone, reflecting the difficulty of achieving either ultra-hard or ultra-soft properties in opaque mixtures. Moreover, the lower portion of the FPS presents an irregular triangular shape, suggesting that the generalizability of $Y_1$ is more constrained when $Y_2$ falls below Shore 00-40. Further insights into FPS can be obtained by analyzing the roles of $x_1$, $x_2$, and $x_3$ individually.

For FPS(AC) (Fig.~\ref{feasible space} (A)), its primary influence is observed in the upper half of the FPS, where higher $x_1$ values correspond to higher $Y_2$. Notably, the full $Y_1$ range (5-80\%) can be covered when $x_1$ is between 40-50\%. By contrast, FPS(TM) (Fig.~\ref{feasible space} (B)) mainly governs the lower FPS region, where larger $x_2$ values lead to softer $Y_2$. Additionally, the gradients of both $x_1$ and $x_2$ align nearly parallel to $Y_1$, indicating that their variations contribute less significantly to transparency. Finally, FPS(GM) (Fig.~\ref{feasible space} (C)) dominates the left side of the FPS, where $x_3$ remains below 15\% whenever $Y_1 > 40\%$, confirming that GM strongly induces opacity.

\subsection{Inverse Design for Property Tailoring}


The inverse design method for tailoring elastomer properties is illustrated in Fig.~\ref{plot:inverse design method} and described as follows.

\begin{figure*}[!htbp]
	\centering
	\includegraphics[width = 1\hsize]{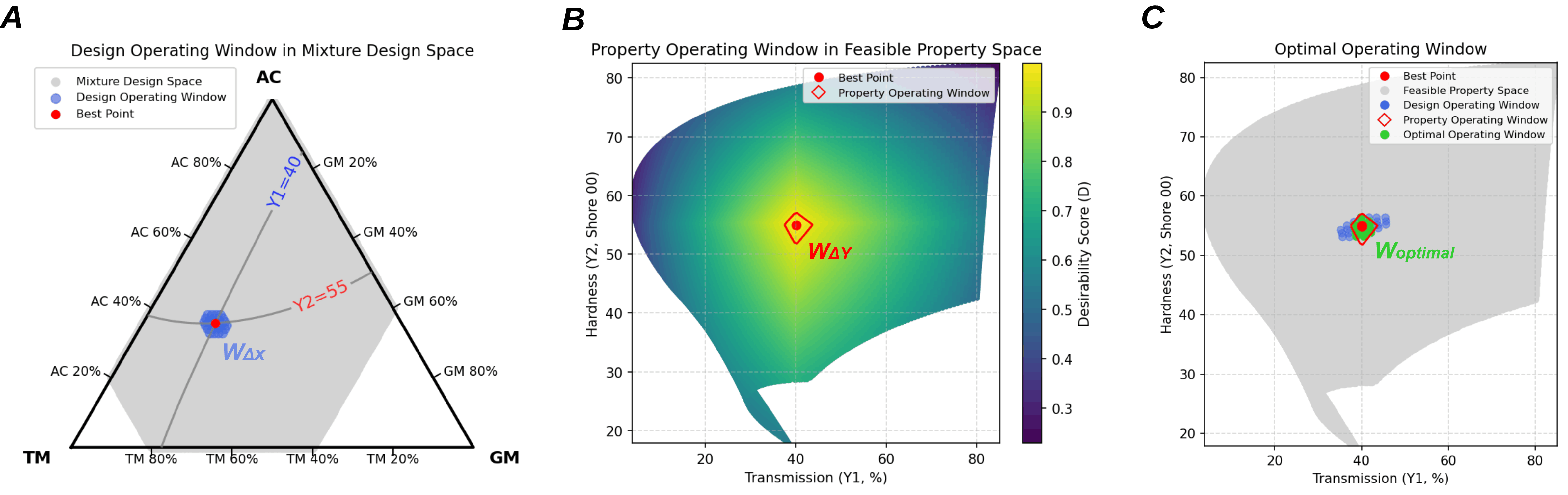}
	\caption{\small A: Design operating window $W_{\Delta x}$ with user-defined $\pm \Delta x\%$. B: Property operating window $W_{\Delta Y}$ with user-defined $- \Delta Y\%$. C: Optimal operating window $W_{optimal}$, defines
    a set of $i$ integer-percentage designs 
 \{$(x_1, x_2, x_3)_{0},...,(x_1, x_2, x_3)_{i}$\} for subsequent printing. } 
	\label{optimal window}
\end{figure*}


\subsubsection{Desirability Function}

Given the property targets of both ($T_{Y_1}$, $T_{Y_2}$), desirability function approach~\cite{costa2011desirability} is employed to achieve the target tunability via multi-objective optimisation to locate \textbf{optimal mixture design} of ($\hat{x}_1, \hat{x}_2, \hat{x}_3$). It evaluates both responses through a common scale of desirability, where three criteria are typically used: (1) Nominal-The-Better (NTB), (2) Larger-The-Better (LTB), and (3) Smaller-The-Better (STB). Their mathematical formulations are given as follows:

\begin{equation}
    d_{NTB}(Y_i) =
    \begin{cases}
    0, & Y_i < L_{Y_i} \ \text{or} \ Y_i > U_{Y_i}, \\
    \left( \dfrac{Y_i - L_{Y_i}}{T_{Y_i} - L_{Y_i}} \right)^r, & L_{Y_i} \leq Y_i \leq T_{Y_i}, \\
    \left( \dfrac{U_{Y_i} - Y_i}{U_{Y_i} - T_{Y_i}} \right)^r, & T_{Y_i} < Y_i \leq U_{Y_i},
    \end{cases}
\end{equation}

\begin{equation}
    d_{LTB}(Y_i) =
    \begin{cases}
    0, & Y_i < L_{Y_i}, \\
    \left( \dfrac{Y_i - L_{Y_i}}{U_{Y_i} - L_{Y_i}} \right)^r, & L_{Y_i} \leq Y_i \leq U_{Y_i}, \\
    1, & Y_i > U_{Y_i},
    \end{cases}
\end{equation}

\begin{equation}
    d_{STB}(Y_i) =
    \begin{cases}
    0, & Y_i > U_{Y_i}, \\
    \left( \dfrac{Y_i - U_{Y_i}}{L_{Y_i} - U_{Y_i}} \right)^r, & L_{Y_i} \leq Y_i \leq U_{Y_i}, \\
    1, & Y_i < L_{Y_i}.
    \end{cases}
\end{equation}

For NTB, the objective is to bring the response value $Y_i$ as close as possible to a predefined target $T_{Y_i}$, with desirability $d_{NTB}(Y_i)$ decreasing as the response deviates from this target. In contrast, LTB and STB impose one-sided preferences: LTB aims to maximize the response value $Y_i$ (towards an upper bound $U_{Y_i}$), while STB aims to minimize it (towards a lower bound $L_{Y_i}$). To facilitate intuitive inverse design, a comprehensive guideline for selecting desirability criteria is summarized in Table~\ref{desirability criteria guidance}. In total, nine criteria guidance are provided to accommodate diverse design objectives, covering a broad spectrum of elastomer property requirements. Specifically, when users aim for a well-defined specification on both transparency and hardness, the NTB/NTB (Id-1) is recommended, as it identifies a crossover point that simultaneously satisfies both targets. If the design goal is to maximize one property while constraining the other to a fixed indicator, then NTB/LTB or NTB/STB (Id-2/3/4/7) are suitable choices. Finally, LTB/STB (Id-5/6/8/9) are most appropriate for users seeking the property boundaries of mixture design. Once the desirability functions ($d_1$, $d_2$) are selected as ($T_{Y_1}$, $T_{Y_2}$), respectively, the overall desirability function $D$~\cite{costa2011desirability} can be formulated as:

\begin{equation}
D(T_{Y_1}, T_{Y_2}) = D(d_1(Y_1), d_2(Y_2)) = 
\left( \prod_{i=1}^{2} d_i(Y_i)^{w_i} \right)^{1/ \sum w_i}
\end{equation}

Where $w_i \in [w_1,w_2]$, denote user-defined weights assigned to criterion of ($d_1$,$d_2$), allowing preference adjustment between ($T_{Y_1}$, $T_{Y_2}$) in cases of objective optimization conflict. In this way, independent targets $T_{Y_i}$ are unified under a single formulation $D$, parameterized by user-defined criteria $d_i$ and weights $w_i$. Then optimization task becomes finding the maximum value of $D$, which yields the optimal composition ($\hat{x}_1, \hat{x}_2, \hat{x}_3$)~\cite{costa2011desirability}:

\begin{equation}
\begin{aligned}
\hat{x}_1, \hat{x}_2, \hat{x}_3 
&= \arg\max_{x_1,x_2,x_3} D(T_{Y_1}, T_{Y_2}) \\
&= \arg\max 
\left( \prod_{i=1}^{2} d_i(Y_i(x_1, x_2, x_3))^{w_i} \right)^{\tfrac{1}{\sum w_i}} \\
\text{s.t.} \quad 
& \sum x_i = 1, \sum w_i = 1\\
& 0 \leq x_1 \leq 1, 
 0 \leq x_2 \leq 0.8, 
 0 \leq x_3 \leq 0.6.
\end{aligned}
\end{equation}

\subsubsection{Optimal Operating Window}

Although the desirability function enables multi-objective optimization, its single optimal solution 
($\hat{x}_1, \hat{x}_2, \hat{x}_3$) is generally obtained as floating-point percentages. Such fractional 
compositions cannot be directly implemented on PolyJet printer via digital material assignment for monolithic manufacturing. To enhance practical 
feasibility, it is therefore necessary to explore a set of near-optimal solutions with integer-valued 
compositions. For this purpose, we introduce the optimal operating window approach.  

As illustrated in Fig.~\ref{optimal window} (A), the first step is to locate the optimal solution 
($\hat{x}_1, \hat{x}_2, \hat{x}_3$) in the mixture design space. Around this point, floating thresholds 
($\pm \Delta x\%$) are applied to each component, thereby generating a \textbf{design operating window} 
$W_{\Delta x}$. In parallel, within the feasible property space (Fig.~\ref{optimal window} (B)), 
the corresponding ReSM predictions ($\hat{Y_1}, \hat{Y_2}$) of ($\hat{x}_1, \hat{x}_2, \hat{x}_3$) are taken as 
benchmarks, where a tolerance margin ($-\Delta Y\%$) is then applied along its descending gradient of the 
desirability function $\Delta D$, yielding a \textbf{property operating window} $W_{\Delta Y}$. Finally, the intersection of $W_{\Delta x}$ and $W_{\Delta Y}$ defines \textbf{optimal operating window} $W_{optimal}$ in feasible 
property space. As shown in Fig.~\ref{optimal window} (C), $W_{optimal}$ contains $i$ set of 
integer-percentage designs $\{(\hat{x}_1,\hat{x}_2,\hat{x}_3)_0,\ldots,(\hat{x}_1,\hat{x}_2,\hat{x}_3)_i\}$, which can be then employed for monolithic 
manufacturing.

\subsubsection{Inverse Design Error Analysis}

To evaluate the proposed method in Fig.~\ref{plot:inverse design method}, we introduce the concept of the \textbf{inverse design error}, which is composed of two components: 

\begin{enumerate}
    \item \textbf{Error1}: defined as the discrepancy between the desired elastomer property targets $(T_{Y_1}, T_{Y_2})$ and the corresponding property prediction from ReSM $\{(\hat{Y}_1,\hat{Y}_2)_0,\ldots,(\hat{Y}_1,\hat{Y}_2)_i\}$, based on optimal operating window $\{(\hat{x}_1,\hat{x}_2,\hat{x}_3)_0,\ldots,(\hat{x}_1,\hat{x}_2,\hat{x}_3)_i\}$.
    
    \item \textbf{Error2}: defined as the deviation between the ReSM-predicted properties $\{(\hat{Y}_1,\hat{Y}_2)_0,\ldots,(\hat{Y}_1,\hat{Y}_2)_i\}$, and the experimentally evaluated properties of the 3D-printed elastomer $\{(R_{Y_1},R_{Y_2})_0,\ldots,(R_{Y_1},R_{Y_2})_i\}$.
\end{enumerate}

In terms of target tunability, Error1 assesses the known error of the multi-objective optimisation based on the desirability function when a property tailoring target is specified (Fig.~\ref{inverse design pipeline} (B)). This enables an evaluation of whether the chosen desirability criteria and weight parameters lead to an optimal mixture design consistent with user-defined preferences. By contrast, Error2 serves to evaluate the prediction uncertainty of the ReSM in material formulation (Fig.~\ref{inverse design pipeline} (A)). In this case, for a given mixture design, the greater the predictive accuracy of the ReSM, the closer the realised elastomer properties will be to the predicted values, thereby reducing Error2.

\begin{figure*}[!htbp]
	\centering
	\includegraphics[width = 1\hsize]{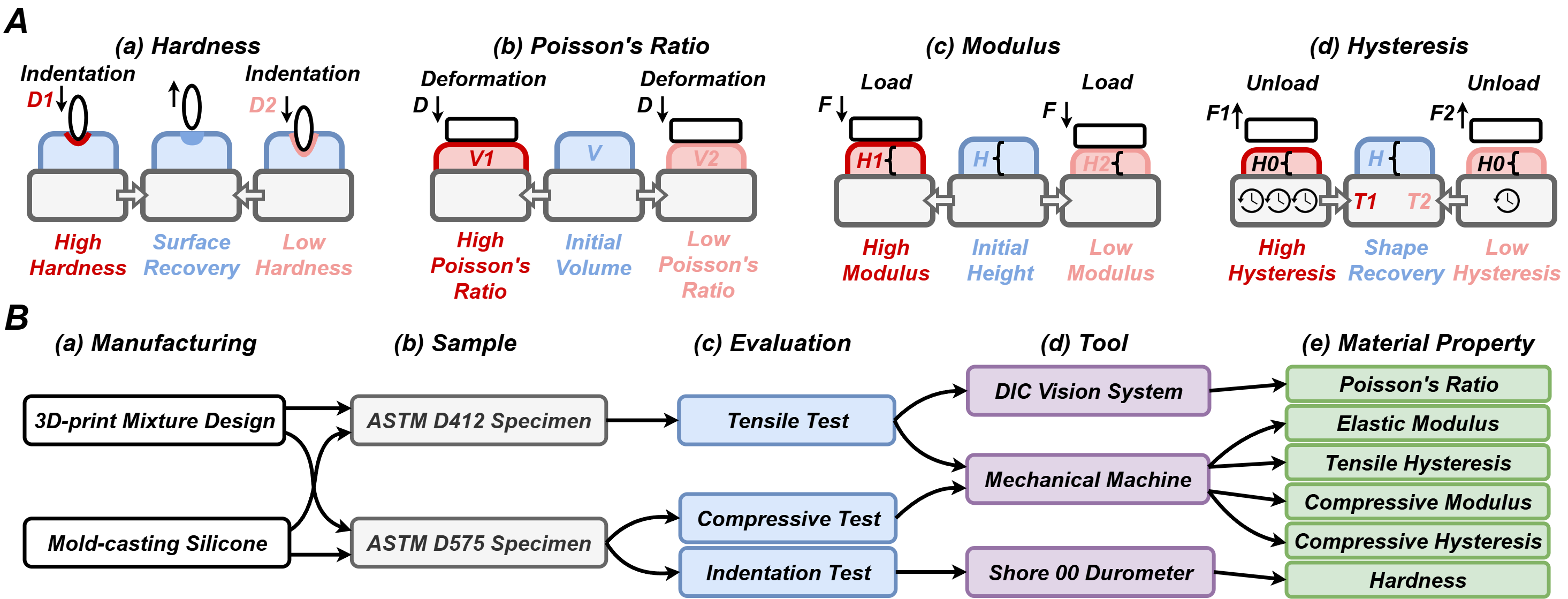}
	\caption{A: Influence of elastomer material properties on tactile sensing. (a) Hardness reflects the resistance of the sensor surface to indentation; (b) Poisson's ratio characterises the volumetric change of the elastomer during deformation; (c) Modulus indicates the resistance to bulk deformation under external loading; (d) Hysteresis quantifies the recovery rate of the sensor after load removal. 
    B: Investigation pipeline for elastomer material characterisation. (a) i-Tac mixture design and mold-casting methods; (b) Sample preparation following ASTM standards; (c) Three mechanical evaluation tests; (d) Instrumentation for experimental data acquisition; (e) Six material properties considered in the analysis.}
	\label{material property setup}
\end{figure*}

\begin{figure*}[!htbp]
	\centering
	\includegraphics[width = 1\hsize]{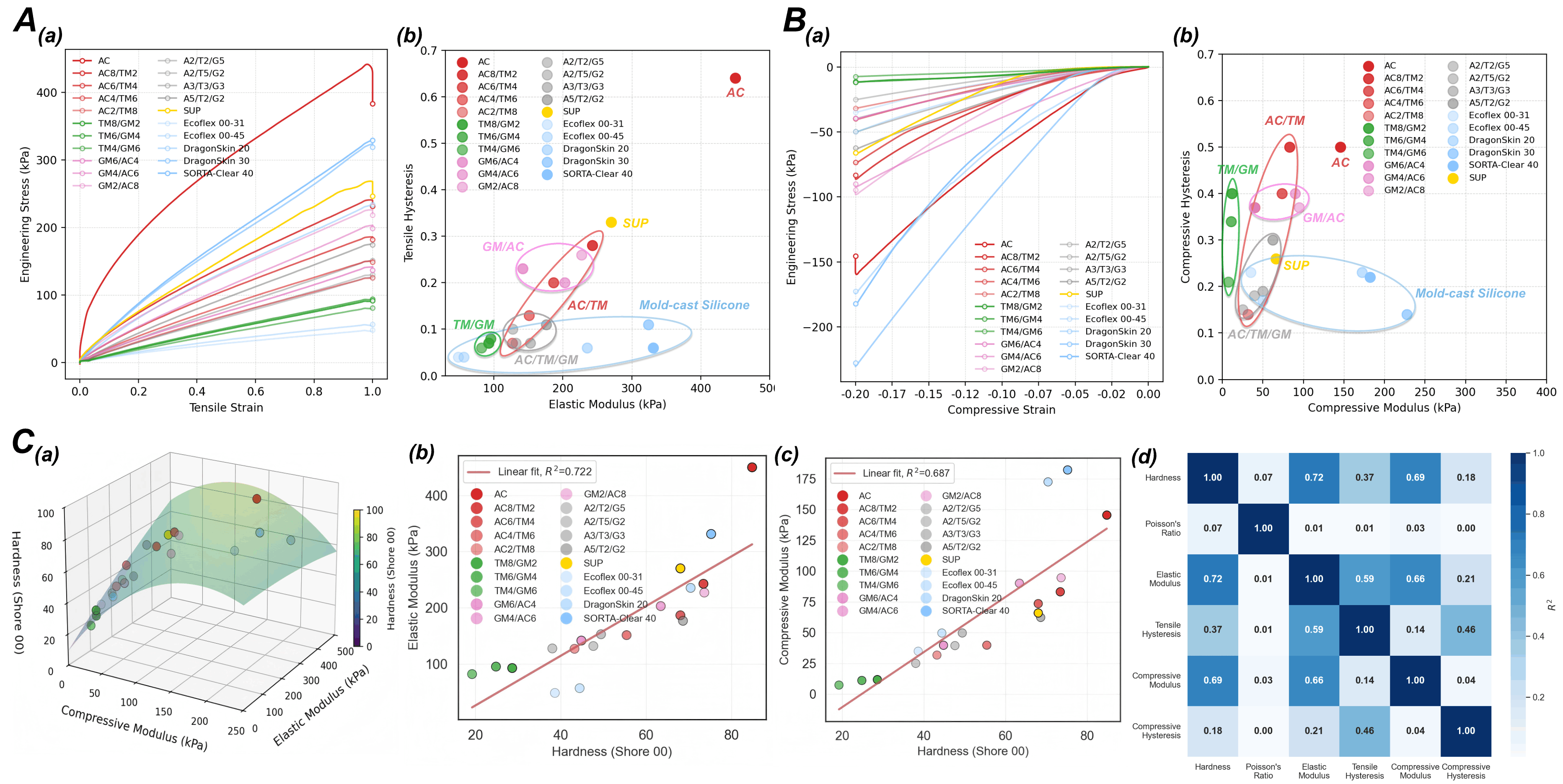}
	\caption{Material property evaluation results. A: Uniaxial tensile testing results. (a) Representative stress-strain curve under uniaxial tension; (b) Relationship between tensile hysteresis and elastic modulus. B: Uniaxial compression testing results. (a) Representative stress-strain curve under uniaxial compression; (b) Relationship between compressive hysteresis and compressive modulus. C: Correlations among elastomer material properties. (a) Response surface fitting of hardness, elastic modulus, and compressive modulus; (b,c) Linear relationships between elastomer hardness and elastic/compressive modulus; (d) Cross-property correlations.
}
	\label{material property fig}
\end{figure*}

\section{Experiment}

As illustrated in Fig.~\ref{inverse design pipeline}, the proposed i-Tac pipeline provides three principal advantages: 
(1) it introduces an AC/TM/GM mixture design strategy to enhance the property scalability of 3D-printed elastomers; 
(2) it establishes an inverse design framework for realizing target tunability of elastomer property; and 
(3) the PolyJet-based monolithic manufacturing approach enables direct fabrication of complete VBTS contact modules, simplifying the production workflow and improving manufacturing efficiency. 
In this section, a series of experiments are conducted to validate the aforementioned advantages.

\subsection{Material Property Evaluation of Mixture Designs}

\subsubsection{Material Properties for Tactile Sensing}

As shown in Fig.~\ref{material property setup}(A), multiple material properties of VBTS elastomers critically influence tactile sensing performance: 
(1) hardness characterises the resistance of an elastomer to surface indentation; 
(2) Poisson's ratio quantifies the volumetric change during deformation; 
(3) the elastic modulus reflects the resistance to bulk deformation under tensile or compressive loading; and 
(4) hysteresis describes the recovery behaviour when the external load is removed. 
To systematically analyse these factors, we establish the \textbf{material property evaluation pipeline} illustrated in Fig.~\ref{material property setup}(B). This framework enables comparative property assessment among 3D-printed mixture designs and in comparison with mold-cast silicone elastomers.

For benchmarking purposes, five commercial silicone products widely adopted in VBTS research~\cite{zhang2022hardware} are included, covering a broad spectrum from transparent to opaque and from ultra-soft to relatively stiff materials:  
(1) \textbf{Ecoflex 00-31 and 00-45}\footnote{\url{https://www.smooth-on.com/product-line/ecoflex/}}: ultra-soft, platinum-cured silicones with excellent optical clarity, curing at room temperature within approximately 4~hrs, with Shore 00-31 and 00-45 hardness, respectively;  
(2) \textbf{DragonSkin 20 and 30}\footnote{\url{https://www.smooth-on.com/product-line/dragon-skin/}}: opaque, platinum-cured silicone rubbers with high tensile strength, curing at room temperature within 4-16~hrs, with Shore 20A and 30A hardness;  
(3) \textbf{SORTA-Clear 40}\footnote{\url{https://www.smooth-on.com/product-line/sorta-clear/}}: a near-transparent, platinum-cured silicone rubber, curing at room temperature within approximately 16~hrs, with Shore 40A hardness.

The results of the tensile and compression evaluations are presented in Fig.~\ref{material property fig}(A,B), with detailed experimental procedures provided in Appendix~A. 
For tensile behaviour, commercial silicones typically exhibit elastic moduli in the range of 50-350~kPa. In contrast, conventional PolyJet elastomers such as AC and SUP show comparatively high stiffness (450~kPa and 270~kPa, respectively). 
By employing the proposed mixture design, the effective modulus range of 3D-printed elastomers is expanded to approximately 100-250~kPa. Meanwhile, tensile hysteresis is substantially reduced from 0.3-0.6 to 0.05-0.3, approaching the 0.05-0.1 range commonly observed in commercial silicones. A similar trend is observed under compression. The mixture designs extend the compressive modulus range from the baseline values of AC (150~kPa) and SUP (65~kPa) to nearly 10-100~kPa, while reducing compression hysteresis from 0.5 and 0.3 to 0.1-0.5. These values move closer to the typical ranges of commercial silicones (compressive modulus: 30-220~kPa; hysteresis: 0.1-0.3).

Overall, compared to typical PolyJet-printed elastomers, the material property evaluation results indicate that AC/TM/GM mixture designs provide: 
(i) improved force sensitivity due to reduced modulus, 
(ii) enhanced flexibility in sensor design enabled by an expanded modulus range, and 
(iii) faster dynamic response associated with lower hysteresis. 
Collectively, these improvements help narrow the performance gap between PolyJet-printed elastomers and conventional mold-cast silicones. 
A comprehensive summary of the evaluated material properties is provided in Table~\ref{material property compare table}. In addition to hardness, the measured Poisson's ratio of the mixture-designed elastomers fluctuates around 0.4. Considering the measurement uncertainty of the DIC system and the commonly assumed value of 0.49 for nearly incompressible commercial silicones, the volumetric compressibility of the proposed mixture designs is expected to be comparable to that of conventional silicone materials.

\begin{table*}[!htbp]
\caption{Material Property Comparison between 3D-Print Mixture Design and Mold-cast Elastomers}
\centering
\small
\resizebox{\textwidth}{!}{%
\begin{tabular}{|c|c|c|c|c|c|c|}
\hline
Mixture Design & Hardness (Shore 00)  & Poisson's Ratio & Elastic Modulus (kPa) & Tensile Hysteresis & Compressive Modulus (kPa) & Compressive Hysteresis \\ \hline
AC10          & 84.8          & 0.42            & 449.87                 & 0.64             & 145.45              & 0.50 \\ \hline
AC8/TM2        & 73.4         & 0.38            & 242.68                 & 0.28             & 83.15               & 0.50 \\ \hline
AC6/TM4          & 68.0       & 0.41            & 186.60                 & 0.20             & 73.55               & 0.40 \\ \hline
AC4/TM6        & 55.4         & 0.42            & 151.35                 & 0.13             & 39.77               & 0.37 \\ \hline
AC2/TM8        & 43.2         & 0.41            & 127.02                 & 0.07             & 31.75               & 0.14 \\ \hline
TM8/GM2       & 28.6         & 0.40            & 92.80                  & 0.07             & 11.83               & 0.40 \\ \hline
TM6/GM4      & 24.8           & 0.45            & 95.39                  & 0.08             & 11.13               & 0.34 \\ \hline
TM4/GM6        & 19.2         & 0.45            & 82.03                  & 0.06             & 7.39                & 0.21 \\ \hline
AC4/GM6       & 44.8          & 0.42            & 141.68                 & 0.23             & 39.70               & 0.37 \\ \hline
AC6/GM4       & 63.4          & 0.36            & 202.99                 & 0.20             & 90.09               & 0.40 \\ \hline
AC8/GM2       & 73.6          & 0.38            & 227.04                 & 0.26             & 94.56               & 0.37  \\ \hline
A3/T3/G3     & 49.4        & 0.43            & 152.88                 & 0.07             & 49.71               & 0.19 \\ \hline
A5/T2/G2     & 68.6        & 0.37            & 176.83                 & 0.11             & 62.32               & 0.30 \\ \hline
A2/T5/G2    & 47.6         & 0.39            & 131.95                 & 0.07             & 39.43               & 0.18 \\ \hline
A2/T2/G5      & 38.0       & 0.45            & 127.68                 & 0.10             & 24.98               & 0.15 \\ \hline
SUP  & 68.0   & 0.43            & 270.08                 & 0.33             & 65.95               & 0.26  \\ \hline
Ecoflex 00-31     & 38.6     & 0.35            & 48.53                 & 0.04             & 34.89              & 0.23 \\ \hline
Ecoflex 00-45    & 44.4      & 0.33            & 56.99                 & 0.04             & 49.67              & 0.17 \\ \hline
DragonSkin 20     & 70.4     & 0.41            & 235.40                 & 0.06             & 172.45              & 0.23 \\ \hline
DragonSkin 30     & 75.0     & 0.38            & 324.30                 & 0.11             & 227.77              & 0.14 \\ \hline
SORTA-Clear 40      & 75.2    & 0.40            & 331.25                 & 0.06             & 182.16              & 0.22 \\ \hline
\end{tabular}%
}
\label{material property compare table}
\end{table*}


\subsubsection{Material Property Correlation}

Based on the above investigation, both 3D-print mixture designs and mold-cast silicone have considerable overlap in material property. Therefore, another investigation is introduced to analyse the correlation between their material properties. 

From Fig.~\ref{material property fig} (C.a), hardness, elastic modulus, and compressive modulus of all materials can be fitted to a response surface. If exclude outliers, the most material points converge into a narrow inclined plane, which implies existing correlations. Linear regression model is used to analyse this trend, which indicates a positive correlation between hardness and both modulus (Fig.~\ref{material property fig} (C.b/c)). More comprehensively, all correlations across material properties are analysed with $R^2$ results shown in Fig.~\ref{material property fig} (C.d). In summary, hardness and modulus are closely connected with $R^2$ locates around 0.7, while Poisson's ratio is not significantly correlated with others. Similar to modulus, correlation also exists between tensile and compressive hysteresis with $R^2$ locates around 0.5.

The above analyses are essential because, in many cases, a complete characterisation of material properties cannot be achieved due to experimental constraints or application-specific conditions. In such situations, rational correlations can be employed to predict unknown properties, for example, by estimating the elastic modulus from the more readily measured elastomer hardness. The observed correlations among material properties across mixture designs further demonstrate the scalability and tunability of the i-Tac framework in inverse design tasks. Specifically, if i-Tac can tailor elastomer hardness, it can be readily extended to achieve comparable performance in targeting elastic or compressive modulus, thereby broadening the applicability of the framework.

\subsection{Inverse Design Performance on Property Tailoring}

\begin{figure*}[!htbp]
	\centering
	\includegraphics[width = 1\hsize]{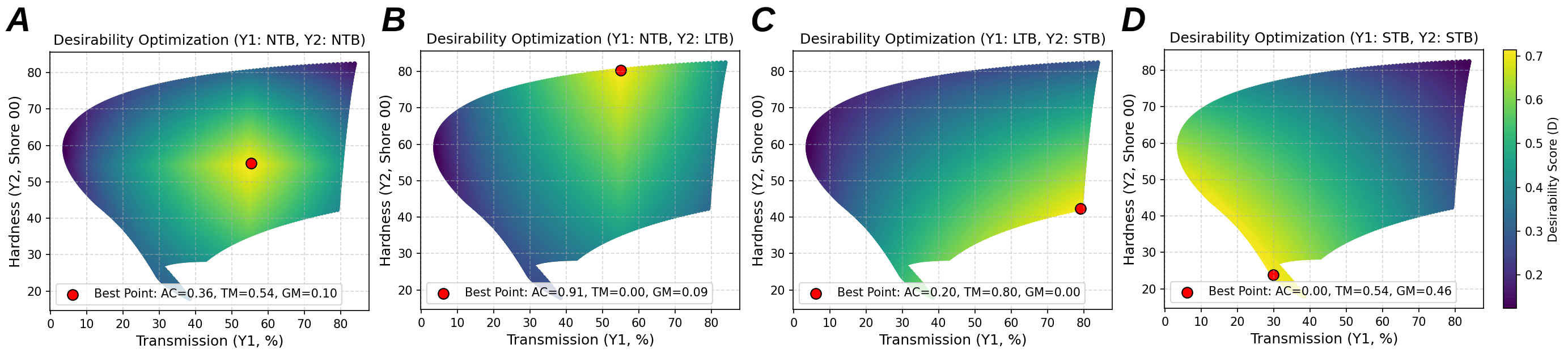}
	\caption{\small Case study of desirability optimization in terms of criteria guideline. A: NTB/NTB; B: NTB/LTB; C: LTB/STB; D: STB/STB.} 
	\label{desirability inverse example}
\end{figure*}

\subsubsection{Tailoring Target from Desirability Criteria Guideline}

To assess the effectiveness of the proposed desirability function approach for inverse design, nine groups of desired elastomer properties $(T_{Y_1}, T_{Y_2})$ were defined according to \textbf{desirability criteria guideline} in Table~\ref{desirability criteria guidance}. The corresponding evaluation results are summarised in Fig.~\ref{desirability inverse example} and Table~\ref{inverse design example table}. 

For Id-1, Id-2, and Id-3, the $T_{Y_1}$ is consistently set to 55\% under the NTB criterion. Following multi-objective optimisation using the desirability function, although $T_{Y_2}$ criterion varies to LTB or STB, their predicted values $\hat{Y}_1$ based on optimal mixture design $(\hat{x_1}, \hat{x_2}, \hat{x_3})$ remain close to the desired 55\%. Similarly, Id-4, Id-5, and Id-6 all adopt the LTB criterion for $T_{Y_1}$, while Id-7, Id-8, and Id-9 adopt the STB criterion. Their corresponding predictions $\hat{Y}_1$ successfully reproduce the expected trends, ranging correctly from higher values (79.18\%-83.91\%) to lower values (17.5\%-29.83\%) in terms of selected $T_{Y_1}$ of LTB and STB, respectively.

\begin{figure*}[!htbp]
	\centering
	\includegraphics[width = 1\hsize]{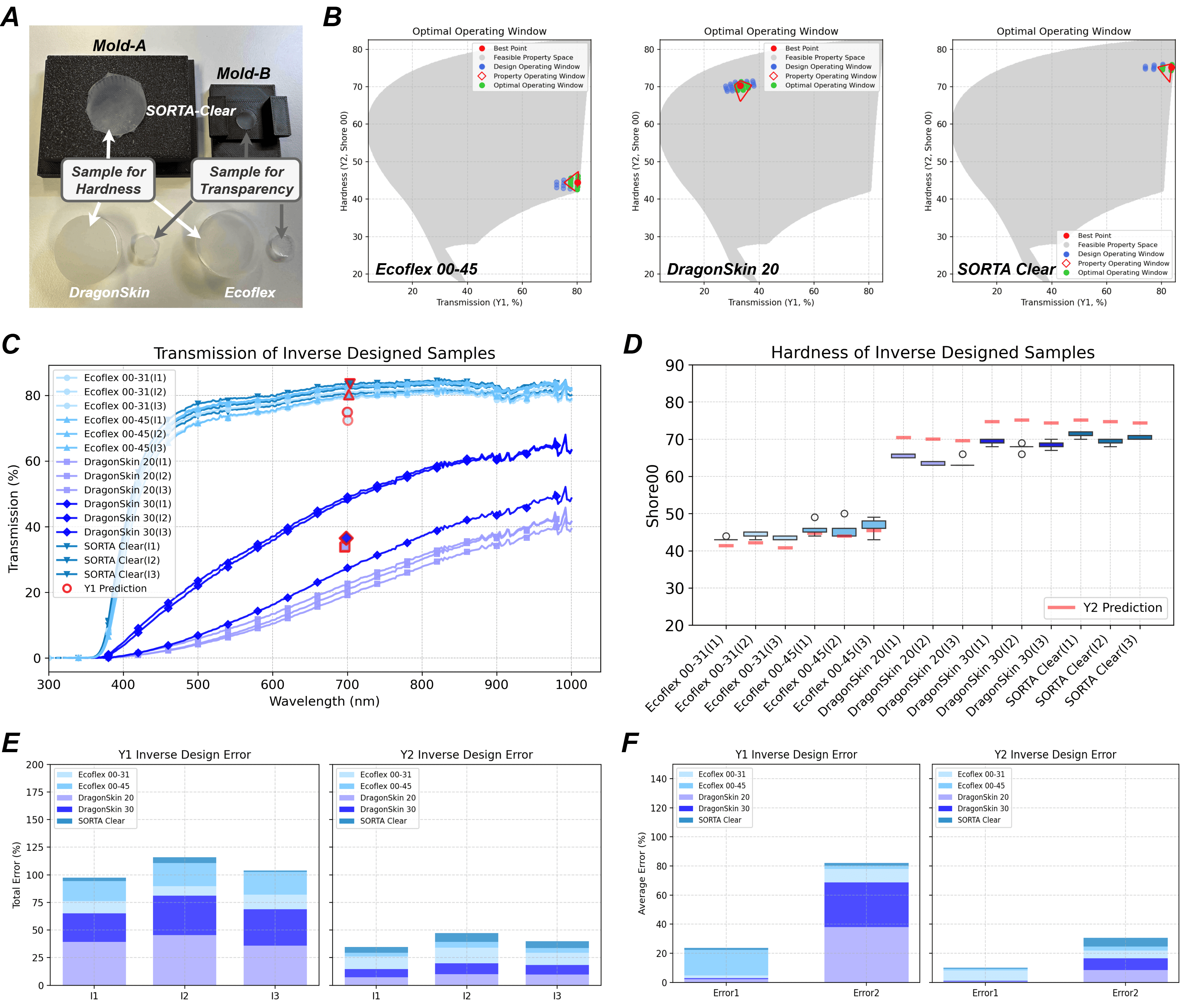}
	\caption{\small A: Mold-casting silicone (Ecoflex, DragonSkin, SORTA Clear). B: Optimal operating window tailoring from real silicone. C: Real transparency $R_{Y_1}$ and predicted $\hat{Y}_1$ (red edges) of inverse design samples. D: Real hardness $R_{Y_2}$ and predicted $\hat{Y}_2$ (red lines) of inverse design samples. E: Accumulated inverse design error compared between $I_1$/$I_2$/$I_3$. F: Average inverse design error compared between Error1/Error2.} 
	\label{real silicone inverse design experiment}
\end{figure*}

Furthermore, in Id-1 through Id-7, the weights $w_1$/$w_2$ are set to 0.5:0.5, thereby ensuring no objective bias in the desirability optimisation. For instance, Id-1 specifies both targets $T_{Y_1}$ and $T_{Y_2}$ as 55, yielding predicted values of $\hat{Y}_1 = 55.3\%$ and $\hat{Y}_2 = 55.08\%$. In contrast, Id-8 and Id-9 share the same $T_{Y_1}$ but adopt different weighting schemes. Specifically, Id-8, with $w_1$/$w_2 = 3:7$, is more strongly biased towards $T_{Y_2}$ (hardness), whereas Id-9, with $w_1$/$w_2 = 6:4$, places greater emphasis on $T_{Y_1}$ (transparency). It further indicates that the optimal inverse design $(\hat{x}_1,\hat{x}_2,\hat{x}_3)$ for Id-8 is $(74,0,26)$, involving only $x_1$ and $x_3$ while excluding $x_2$, since the AC/GM mixture can simultaneously provide moderate opacity and enhanced hardness. By comparison, the optimal mixture design for Id-9 is $(0,54,46)$, which relies exclusively on $x_2$ and $x_3$ rather than $x_1$, in order to achieve optimal opacity and moderate softness. This is consistent with the fact that AC-based mixtures inherently increase both transparency and rigidity found during material formulation (Fig.~\ref{ReSM_2d}).


These results demonstrate that the proposed desirability function method is capable of faithfully realising diverse inverse design tasks defined by the criteria guidelines, while also allowing user preferences to be flexibly incorporated into above multi-objective optimisation through adjustable weight coefficients.

\begin{table}[!htbp]
    \centering
    \footnotesize
    \caption{Inverse Design of Desirability Criteria Guideline}
    \begin{tabular}{|c|c|c|c|c|c|c|c|c|c|}
    \hline
         Id & \textbf{$T_{Y_1}$} & \textbf{$T_{Y_2}$} & \textbf{$w_1$} & \textbf{$w_2$} & \textbf{$\hat{x}_1$} & \textbf{$\hat{x}_2$} & \textbf{$\hat{x}_3$} & \textbf{$\hat{Y}_1$} & \textbf{$\hat{Y}_2$}  \\ \hline
         Id-1 & 55 &  55 & 0.5 & 0.5 & 36 & 54 & 10 & 55.30 & 55.08 \\ \hline
         Id-2 & 55 & LTB & 0.5 & 0.5 & 91 & 0 & 9 & 54.98 & 80.41 \\ \hline
         Id-3 & 55 & STB & 0.5 & 0.5 & 9 & 80 & 11 & 55.04 & 35.43 \\ \hline
         Id-4 & LTB & 60 & 0.5 & 0.5 & 46 & 54 & 0 & 80.73 & 60.01 \\ \hline
         Id-5 & LTB & LTB & 0.5 & 0.5 & 100 & 0 & 0 & 83.91 & 82.49 \\ \hline
         Id-6 & LTB & STB & 0.5 & 0.5 & 20 & 80 & 0 & 79.18 & 42.33 \\ \hline
         Id-7 & STB & 60 & 0.5 & 0.5 & 33 & 7 & 60 & 17.5 & 40.15 \\ \hline
         Id-8 & STB & LTB & 0.3 & 0.7 & 74 & 0 & 26 & 17.98 & 72.91 \\ \hline
         Id-9 & STB & STB & 0.6 & 0.4 & 0 & 54 & 46 & 29.83 & 23.89 \\ \hline
    \end{tabular}
    \label{inverse design example table}
\end{table}

\subsubsection{Tailoring Target from Real Silicone Samples}

\begin{table}[!htbp]
    \centering
    \caption{Inverse Design of Real Silicone Samples}
    \footnotesize
    \setlength{\tabcolsep}{4pt}
    \renewcommand{\arraystretch}{1.1}
    \begin{tabular}{|c|c|c|c|c|c|c|c|}
    \hline
        \textbf{Real Sample} & $T_{Y_1}$ & $T_{Y_2}$ & $\hat{x}_1$ & $\hat{x}_2$ & $\hat{x}_3$ & \textbf{Desirability} & \textbf{Design} \\ \hline

        \multirow{3}{*}{Ecoflex 00-31} 
        & \multirow{3}{*}{\makecell{75.43 \\ (NTB)}}
        & \multirow{3}{*}{\makecell{38.60 \\ (STB)}}
        & 18 & 80 & 2 & 0.7322 & $I_1$ \\ \cline{4-8}
        & & & 19 & 79 & 2 & 0.7266 & $I_2$ \\ \cline{4-8}
        & & & 17 & 80 & 3 & 0.7243 & $I_3$ \\ \hline

        \multirow{3}{*}{Ecoflex 00-45} 
        & \multirow{3}{*}{\makecell{97.23 \\ (LTB)}}
        & \multirow{3}{*}{\makecell{44.40 \\ (NTB)}}
        & 23 & 77 & 0 & 0.8925 & $I_1$ \\ \cline{4-8}
        & & & 22 & 78 & 0 & 0.8914 & $I_2$ \\ \cline{4-8}
        & & & 24 & 76 & 0 & 0.8856 & $I_3$ \\ \hline

        \multirow{3}{*}{DragonSkin 20} 
        & \multirow{3}{*}{\makecell{33.13 \\ (NTB)}}
        & \multirow{3}{*}{\makecell{70.40 \\ (NTB)}}
        & 65 & 16 & 19 & 0.9932 & $I_1$ \\ \cline{4-8}
        & & & 64 & 17 & 19 & 0.9923 & $I_2$ \\ \cline{4-8}
        & & & 63 & 18 & 19 & 0.9879 & $I_3$ \\ \hline

        \multirow{3}{*}{DragonSkin 30} 
        & \multirow{3}{*}{\makecell{36.23 \\ (NTB)}}
        & \multirow{3}{*}{\makecell{75.00 \\ (NTB)}}
        & 75 & 8  & 17 & 0.9961 & $I_1$ \\ \cline{4-8}
        & & & 76 & 7  & 17 & 0.9935 & $I_2$ \\ \cline{4-8}
        & & & 74 & 9  & 17 & 0.9925 & $I_3$ \\ \hline

        \multirow{3}{*}{SORTA-Clear 40} 
        & \multirow{3}{*}{\makecell{84.73 \\ (LTB)}}
        & \multirow{3}{*}{\makecell{75.20 \\ (NTB)}}
        & 77 & 23 & 0 & 0.9145 & $I_1$ \\ \cline{4-8}
        & & & 76 & 24 & 0 & 0.9117 & $I_2$ \\ \cline{4-8}
        & & & 75 & 25 & 0 & 0.9089 & $I_3$ \\ \hline
    \end{tabular}
    \label{tab:inverse_design_real_silicone}
\end{table}

\begin{table*}[!htbp]
    \centering
    \caption{Inverse Design Performance with Tailoring Target from Real Silicone Samples}
    \footnotesize
    \setlength{\tabcolsep}{4pt}
    \renewcommand{\arraystretch}{1.1}
    \begin{tabular}{|c|c|c|c|c|c|c|c|c|c|c|}
    \hline
        \textbf{Inverse Design} & \textbf{$T_{Y_1}$} & \textbf{$\hat{Y}_1$} & \textbf{$R_{Y_1}$} & \textbf{$Y_1$ Error1} & \textbf{$Y_1$ Error2} & \textbf{$T_{Y_2}$} & \textbf{$\hat{Y}_2$} & \textbf{$R_{Y_2}$} & \textbf{$Y_2$ Error1} & \textbf{$Y_2$ Error2} \\ \hline
        Ecoflex 00-31 ($I_1$) & 75.43 & 74.90 & 82.60 & 0.53 (0.7\%) & -7.70 (10.3\%) & 38.60 & 41.41 & 43.20 & -2.81 (7.2\%) & -1.79 (4.3\%) \\ \hline
        Ecoflex 00-31 ($I_2$) & 75.43 & 74.94 & 81.03 & 0.49 (0.6\%) & -6.09 (8.1\%) & 38.60 & 42.17 & 44.20 & -3.57 (9.2\%) & -2.03 (4.8\%) \\ \hline
        Ecoflex 00-31 ($I_3$) & 75.43 & 72.44 & 79.25 & 2.99 (4.0\%) & -6.81 (9.4\%) & 38.60 & 40.84 & 43.40 & -2.24 (4.8\%) & -2.56 (6.3\%) \\ \hline
        Ecoflex 00-45 ($I_1$) & 97.23 & 80.25 & 79.81 & 16.98 (17.5\%) & 0.44 (0.5\%) & 44.40 & 44.74 & 45.80 & -0.34 (0.8\%) & -1.06  (2.4\%) \\ \hline
        Ecoflex 00-45 ($I_2$) & 97.23 & 80.18 & 82.95  & 17.05 (17.5\%) & -2.77 (3.5\%) & 44.40 & 44.00 & 46.00  & -0.40 (0.9\%) & -2.00 (4.5\%)  \\ \hline
        Ecoflex 00-45 ($I_3$) & 97.23 & 80.31 & 82.72 & 16.92 (17.4\%) & -2.41 (3.0\%) & 44.40 & 45.47 & 46.40 & -1.07 (2.4\%) & -0.93 (2.0\%) \\ \hline
        DragonSkin 20 ($I_1$) & 33.13 & 33.68 & 21.00 & -0.55 (1.7\%) & 12.68 (37.6\%) & 70.40 & 70.51 & 65.60 & -0.11 (0.2\%) & 4.91 (6.9\%) \\ \hline
        DragonSkin 20 ($I_2$) & 33.13 & 33.84 & 19.14 & -0.71 (2.1\%) & 14.70 (43.4\%) & 70.40 & 70.06 & 63.40 & 0.34 (0.5\%) & 6.66 (9.5\%) \\ \hline
        DragonSkin 20 ($I_3$) & 33.13 & 34.00 & 22.70 & -0.87 (2.6\%) & 11.3 (33.2\%) & 70.40 & 69.62 & 63.60 & 0.78 (1.1\%) & 6.02 (8.6\%) \\ \hline
        DragonSkin 30 ($I_1$) & 36.23 & 36.54 & 27.39 & -0.31 (0.9\%) & 9.15 (25.0\%) & 75.00 & 74.78 & 69.40 & 0.22 (0.3\%) & 5.38 (7.2\%) \\ \hline
        DragonSkin 30 ($I_2$) & 36.23 & 36.40 & 49.14 & -0.17 (0.5\%) & -12.74 (35.0\%) & 75.00 & 75.16 & 67.80  & -0.16 (0.2\%) & 7.36 (9.8\%) \\ \hline
        DragonSkin 30 ($I_3$) & 36.23 & 36.67 & 48.30 & -0.44 (1.2\%) & -11.63 (31.7\%) & 75.00 & 74.40 & 68.60  & -0.60 (0.8\%) & 5.80 (7.8\%) \\ \hline
        SORTA-Clear 40 ($I_1$) & 84.73 & 83.65 & 82.16 & 1.08 (1.3\%) & 1.49 (1.8\%) & 75.20 & 75.18 & 71.20  & 0.02 (0.0\%) & 3.98 (5.3\%) \\ \hline
        SORTA-Clear 40 ($I_2$) & 84.73 & 83.59 & 80.40 & 1.14 (1.3\%) & 3.19 (3.8\%) & 75.20 & 74.78 & 69.40  & 0.42 (0.6\%) & 5.38 (7.2\%) \\ \hline
        SORTA-Clear 40 ($I_3$) & 84.73 & 83.53 & 83.53 & 1.20 (1.4\%) & 0 (0.0\%) & 75.20 & 74.38 & 70.60  & 0.82 (1.1\%) & 3.78 (5.1\%) \\ \hline
    \end{tabular}
    \label{inverse design error table}
\end{table*}

As shown in Fig.~\ref{real silicone inverse design experiment} (A), two casting molds, Mold-A and Mold-B, were fabricated with dimensions identical to the specimens used in mixture experiments. Subsequently, all silicones were prepared and cast at room temperature in accordance with the manufacturers’ guidelines, followed by property testing. The resulting measurements were then used as $T_{Y_1}$ and $T_{Y_2}$, as summarised in Table~\ref{tab:inverse_design_real_silicone}. In this process, most desirability criteria were specified as NTB, with all desirability weights $w_1$/$w_2$ set to 0.5:0.5. In addition, ultra-soft Ecoflex 00-31 was assigned the STB criterion for $T_{Y_2}$ in order to minimise hardness, whereas $T_{Y_1}$ of ultra-clear Ecoflex 00-45 and SORTA Clear were assigned the LTB criterion to maximise transparency. The optimal operating window $\{(\hat{x}_1,\hat{x}_2,\hat{x}_3)_0,\ldots,(\hat{x}_1,\hat{x}_2,\hat{x}_3)_i\}$ was then determined by applying tolerance thresholds of $\Delta x\%$ and $\Delta Y\%$, both set at 3\%, as illustrated in Fig.~\ref{real silicone inverse design experiment} (B). Among these, three inverse designs with the highest desirability values were selected and denoted as $I_1$, $I_2$, and $I_3$, with their corresponding mixture designs summarised in Table~\ref{tab:inverse_design_real_silicone}. Finally, based on the above inverse designs, 15 groups of 3D-printed specimens were fabricated through monolithic manufacturing, each consisting of two optical samples and one mechanical sample (45 specimens in total). Their property evaluation results are presented in Fig.~\ref{real silicone inverse design experiment} (C) and (D).



In Fig.~\ref{real silicone inverse design experiment} (C), the predicted elastomer transparency, $\hat{Y}_1$, are highlighted with red edges. At $\lambda = 700$~nm, the experimentally measured $R_{Y_1}$ values for the real specimens of Ecoflex and SORTA Clear inverse designs closely match their predictions, both falling within the range of 75\%-80\%. For the DragonSkin inverse designs, the $R_{Y_1}$ values of real samples lie between 20\% and 45\%, covering the predicted range of approximately 35\%-40\%. In Fig.~\ref{real silicone inverse design experiment} (D), the predicted elastomer hardness, $\hat{Y}_2$, are indicated by red lines. The 3D-printed elastomers of Ecoflex inverse designs exhibit very low hardness, with measured $R_{Y_2}$ values ranging from Shore 00-40 to 00-50, in close agreement with their $\hat{Y}_2$ predictions. By contrast, the DragonSkin and SORTA Clear inverse designs show a marked increase in hardness towards Shore 00-60 and 00-70, although most of their measured hardness $R_{Y_2}$ are slightly (around 5 units) lower than $\hat{Y}_2$ predictions. These results indicate that the proposed inverse design method achieves a close alignment between predictions and real measurements, particularly for clear or softer silicone systems, while deviations remain in the translucent and harder formulations.

To quantitatively assess the above findings, two types of inverse design error were analysed, following the definitions in Fig.~\ref{plot:inverse design method}. As summarised in Table~\ref{inverse design error table}, \textbf{Error~1}, which represents the difference between the desired target $T_{Y}$ and the ReSM prediction $\hat{Y}$, is generally small. Only Ecoflex~00-31 and Ecoflex~00-45 exhibit notable discrepancies, with $Y_1$ Error~1 of 17\% and $Y_2$ Error~1 of 5-10\%, respectively. By contrast, the predictions $\hat{Y}$ for the other inverse designs are close to their corresponding $T_{Y}$, with most errors below 1\%. This is primarily because the specified $T_{Y_1}$ of Ecoflex~00-45 and $T_{Y_2}$ of Ecoflex~00-31 lie beyond the feasible property space (97.23\%, Shore 00-38.60). Consequently, their desirability criteria were assigned as LTB and STB (Table~\ref{tab:inverse_design_real_silicone}), which inevitably results in a non-negligible gap of Error~1. 

As for \textbf{Error~2}, defined as the difference between the predicted $\hat{Y}$ and the experimentally measured property $R_Y$ of real 3D-printed samples. In terms of elastomer transparency  $Y_1$, Error~2 for Ecoflex~00-45 and SORTA-Clear is low (0-4\%), demonstrating high predictive accuracy for elastomer transparency. Ecoflex~00-31 follows with an Error~2 of 8-10\%, which remains relatively accurate. By contrast, the DragonSkin designs, being more opaque, show substantially higher Error~2 values (25-43\%), indicating that the transparency ReSM $Y_1(x)$ performs less reliably in the opacity region compared with the clarity region. For elastomer hardness $R_{Y_2}$, all five materials exhibit consistently low Error~2 values, generally within 2-10\%. This suggests that the predictive performance of hardness ReSM $Y_2(x)$ is well balanced across the entire FPS.

As shown in Fig.~\ref{real silicone inverse design experiment} (E), the \textbf{accumulated inverse design errors} of each $I_1$, $I_2$, and $I_3$ are compared in terms of each material. Among them, $I_1$ consistently exhibits the smallest total error across both $Y_1$ and $Y_2$, as its desirability is closest to the optimal inverse design (Table~\ref{tab:inverse_design_real_silicone}). In addition, the \textbf{average inverse design error} across samples, combining Error~1 and Error~2, is presented in Fig.~\ref{real silicone inverse design experiment} (F). For both $Y_1$ and $Y_2$, Error~2 is consistently larger than Error~1, indicating that the principal gap in inverse design for tailoring elastomer property lies between property prediction $\hat{Y}$ and real measurement $R_Y$ of 3D-printed elastomer, rather than between the tailoring target $T_Y$ and $\hat{Y}$, which highlights the necessary of material formulation and ReSM optimization. From this analysis, since the largest errors are observed in transparency of DragonSkin inverse designs, it suggests that the transparency ReSM $Y_1(x)$ (Fig.~\ref{ReSM_2d} (A)) requires further refinement in the opacity region.

\begin{figure*}[!htbp]
	\centering
	\includegraphics[width = 1\hsize]{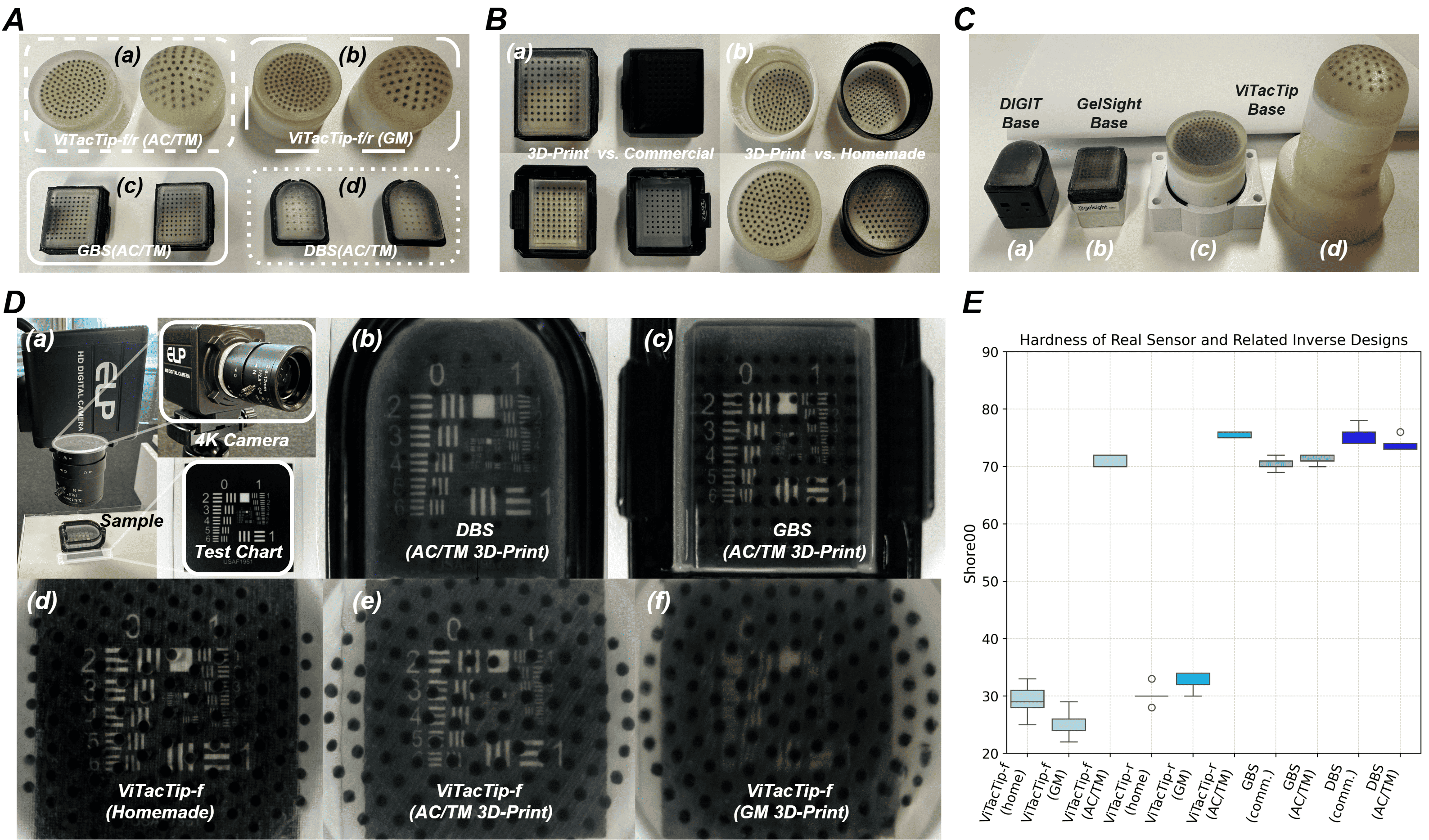}
	\caption{\small A: (a) ViTacTip-f/r with AC/TM elastomer, (b) ViTacTip-f/r with pure GM elastomer, (c) GBS with AC/TM elastomer, (d) DBS with AC/TM elastomer. B: (a) Comparison between 3D-print GBS and commercial GelSight; (b) Comparison between 3D-print and homemade ViTacTip-f. C: VBTS base mounted with 3D-print contact modules, (a) DIGIT base, (b) GelSight base, (c/d) ViTacTip base v1/v2. D: (a) Experiment setup of optical resolution using 4K camera and USAF 1951 test chart. (b/c) Observed images for 3D-print GBS/DBS. (d-f) Observed images for homemade/3D-print ViTacTip-f. E: Hardness comparison between real sensor samples.}
	\label{real_sensor_test}
\end{figure*}

\begin{table*}[!htbp]
    \centering
    \caption{Optical Resolution of Real Sensor Samples}
    \footnotesize
    \setlength{\tabcolsep}{4pt}
    \renewcommand{\arraystretch}{1.1}
    \begin{tabular}{|c|c|c|c|c|c|c|}
    \hline
        \textbf{Parameter} & \textbf{No Sample} & \textbf{ViTacTip-f (homemade)} & \textbf{ViTacTip-f (AC/TM)} & \textbf{ViTacTip-f (GM)} & \textbf{GBS (AC/TM)} & \textbf{DBS (AC/TM)} \\ \hline
        Elastomer Height (mm) & - & 5 & 10 & 10 & 6 & 8 \\ \hline
        Group (G) & 3 & 2 & 1 & 0 & 2 & 2 \\ \hline
        Element (E) & 5 & 3 & 6 & 1 & 3 & 2 \\ \hline
        Optical Resolution (lp/mm) & 12.70  & 5.04 & 3.56 & 1.00 & 5.04 & 4.49 \\ \hline
    \end{tabular}
    \label{real_sensor_usaf1951_table}
\end{table*}

\subsection{Real Sensor Property with Inverse Designed Elastomer}

To further validate the proposed i-Tac inverse design technique, we fabricated the entire contact module of VBTS through proposed i-Tac framework, incorporating the skin, marker, and lens components alongside the inverse-designed elastomers. As illustrated in Fig.~\ref{real_sensor_test}, specimens representing various established VBTS designs have been produced to demonstrate the versatility of the inverse design approach, encompassing a broad range of dimensions, configurations, and elastomer properties.
For this validation, we selected GelSight\cite{yuan2017gelsight} and DIGIT\cite{lambeta2020digit} sensors as target, which represents the most widely-used commercial VBTS products. Additionally, an in-house custom-built ViTacTip \cite{fan2024vitactip} is included. It features a PolyJet-printed transparent skin fabricated from Agilus30 Clear and filled with a silicone-based ultra-soft elastomer. Clearly, the aforementioned sensors cover a comprehensive range of VBTS elastomer characteristics, including geometry, dimensions, and material properties. They therefore serve as an appropriate control group for validating the scalability and tunability of the i-Tac inverse design framework in whole-sensor customisation.

In terms of above control group, four types of i-Tac specimens were fabricated: (1) \textbf{flat-shaped ViTacTip (ViTacTip-f)}: 30mm radius, 10mm elastomer height, 1mm skin thickness; (2) \textbf{round-shaped ViTacTip (ViTacTip-r)}: 30mm radius, 10mm elastomer height, 1mm skin thickness; (3) \textbf{GelSight base sample (GBS)}: 25mm width, 5mm elastomer height, 0.5mm skin thickness, no coating; and (4) \textbf{DIGIT base sample (DBS)}: 20mm width, 8mm elastomer height, 0.5mm skin thickness, no coating.



\subsubsection{Inverse Design Process}

When establishing $T_Y$, it is impractical to measure the optical clarity of a physical sensor using a spectrometer. Thereby, the transparency targets $T_{Y_1}$ are set as LTB criteria to reach maximum elastomer clarity. Then, Shore-00 durometer was employed to assess the hardness of control group via indentation on their sensor surfaces. The measured values are as follows: flat ViTacTip (homemade):
29.20, round ViTacTip (homemade):
30.20, GelSight (commercial): 70.40, and DIGIT (commercial): 76.60 (Shore 00). 

Based on these results, two sets of i-Tac inverse design targets were subsequently formulated: (1) The first target utilises \textbf{pure GM design} (GM10, see Table~\ref{table sample data}) as a gel-like elastomer to realise an ultra-soft silicone in ViTacTip with moderate optical clarity. (2) The second set adopts the same formulation as SORTA-Clear's I3 ($(x_1,x_2,x_3)$=(75,25,0), Table.~\ref{tab:inverse_design_real_silicone}), wherein $T_{Y_1}$ adheres to LTB criterion, and $T_{Y_2}$ targets a Shore 00 hardness of 75.20 (Table.~\ref{inverse design error table}), both values align with the properties of GelSight and DIGIT. Accordingly, \textbf{AC/TM(75/25) mixture design} was selected for the inverse design targeting GelSight and DIGIT. This material was also used to fabricate a ViTacTip for comparison with the pure GM variant. The comparison between resulting i-Tac products alongside commercial and homemade control group are illustrated in Fig.~\ref{real_sensor_test} (A-C).

\subsubsection{Optical Transparency}

In Fig.~\ref{real_sensor_test} (D.a), a 4K camera and a USAF 1951 resolution test chart\footnote{\url{https://en.wikipedia.org/wiki/1951_USAF_resolution_test_chart}} are employed to evaluate the optical clarity. The test chart comprises numerous small target patterns featuring a graded series of precise spatial frequency elements. The smallest distinguishable element under acceptable image contrast conditions provides an estimate of the resolution limit. Each sensor sample is placed directly above the test chart, and the finest resolvable element is identified by its group (G) and element (E) indices based on the image captured by the 4K camera. These indices are used to determine the corresponding spatial frequency in line pairs per millimetre (lp/mm), which serves as a quantitative measure of the optical resolution (R) of the sensor sample:

\begin{equation}
\text{R (lp/mm)} = 2^{\text{G} + (\text{E} - 1)/6}
\end{equation}

In this manner, the optical transparency of above sensors can be quantified by comparing the optical resolution with and without the sample in place. To ensure a fair comparison, only flat-shaped designs were selected (Fig.~\ref{real_sensor_test} (D.b-f)), which can maintain consistent contact with the test chart, thereby ensuring valid imaging with clearly discernible pattern elements. The measured optical resolutions are summarised in Table~\ref{real_sensor_usaf1951_table}. The highest resolution, measured without any sample, is 12.70 lp/mm. As a baseline, the homemade flat ViTacTip achieved a resolution of 5.04 lp/mm. The 3D-printed ViTacTip filled with AC/TM attained 3.56 lp/mm, slightly lower than the baseline, whereas the version with pure GM reached only 1.0 lp/mm. This reduction in clarity for the pure GM sample can be attributed to two main factors: (1) firstly, GM exhibits approximately 20\% lower inherent transparency than the AC/TM mixture; (2) secondly, GM provides inferior structural support for the overlying lens during printing, potentially introducing optical aberrations. These issues may be mitigated through structural optimisations, such as reducing the thickness of the elastomer or the span of the lens. In contrast, sensors fabricated with AC/TM demonstrate superior optical performance. For instance, the 3D-printed GBS and DBS achieved resolutions between 4.49 and 5.04 lp/mm.

\subsubsection{Mechanical Hardness}

As summarised in Fig.~\ref{real_sensor_test} (E), the 3D-printed ViTacTip (GM) exhibit Shore 00 hardness values of
25.40 and
32.40, which are comparable to the handmade ViTacTip (29.20,30.20), and significantly lower than the AC/TM ViTacTip (70.80,75.60). These results indicate that the pure GM material extends the feasible property space of i-Tac inverse design into the ultra-soft range (Shore 00-20 to 00-40), enabling the fabrication of sensor designs that demand high mechanical ductility and shear sensitivity. In addition, other two AC/TM-based i-Tac products exhibit hardness values within the Shore 00-70 to 00-75 range, consistent with the AC/TM ViTacTip. For instance, the AC/TM GBS and DBS show hardness values of
71.20 and
73.80, respectively, aligning closely with their commercial control group of GelSight/DIGIT (70.40,76.60). 

The above evaluation results demonstrates that i-Tac inverse designed product not only construct the entire VBTS contact modules of various design geometry via monolithic manufacturing with high fabrication convenience, but also match the material property of handmade/commercial sensor upon both optical clarity and mechanical hardness with property scalability and target tunability.

\section{Discussion}

Despite the promising performance of the proposed i-Tac framework, several limitations remain. 


First, the current ReSMs are fitted to only 16 mixture-design samples, which may limit modelling accuracy in sparsely sampled regions of the composition space. Future work will explore denser material calibration and data-efficient surrogate models to improve predictive robustness.


Second, the biomimetic inspiration in this work should be interpreted primarily at the conceptual level rather than as a full mechanical replication of human skin. Specifically, the proposed mixture design targets the intrinsic material properties of elastomeric micro-elements (material voxels), capturing the constituent-level property blending that also underlies dermal tissue, where collagen, elastin, and ground substance collectively determine baseline material behaviour. We acknowledge that real skin exhibits pronounced anisotropy, viscoelasticity, and layered organisation arising from higher-order structural arrangements across multiple scales. The present study focuses on the material composition dimension as a foundational step. Importantly, this material-level tunability provides an enabling basis for future structural-mechanical co-design, where anisotropic responses may be achieved through the spatial organisation of mixture-designed voxels. In this context, recent multi-material topology optimisation studies have highlighted the advantages of jointly optimising material distribution and structural layout, rather than treating material selection and geometry as separate design problems. For example, recent work on ultra-soft magneto-active structures \cite{perez2025topology} demonstrates that such optimisation can be further enhanced by explicitly incorporating constitutive descriptions of the underlying coupled physical behaviour into the design process.  More broadly, this points towards a physics-informed co-design paradigm that integrates both material and structural design. Although beyond the scope of this work, we believe this represents a promising direction for future research.

Furthermore, the proposed i-Tac framework is inherently extensible beyond the specific material set used in this study. In particular, the mixture design formulation can naturally evolve alongside ongoing iterations of PolyJet printable materials. As new resins with improved optical or mechanical performance become available, they can be incorporated into the mixture space to progressively expand the feasible material property range and further enhance inverse design scalability. Beyond PolyJet technology, the underlying design philosophy of i-Tac is also applicable to other multi-material fabrication paradigms (e.g., DIW-based printing). In such cases, the framework can be transferred with minimal modification by redefining the material coordinate system in the mixture design space and re-fitting the response surface model using the corresponding experimental data, while preserving the desirability-driven optimisation strategy. This adaptability highlights the long-term scalability and cross-platform potential of the proposed inverse design pipeline for VBTS elastomer engineering.

\section{Conclusion}

This study presents i-Tac, an inverse design methodology for 3D-printed elastomers with tailored optical transparency and hardness, aiming to enhance the monolithic manufacturing of vision-based tactile sensors (VBTS) through improved design flexibility and end-product quality. 

Inspired by the hierarchical structure of human skin, where collagen fibres, elastic fibres, and ground substance jointly determine mechanical behaviour, we propose a material mixture design framework comprising three PolyJet materials: AC, TM, and GM. Within this mixture space, a systematic set of samples is fabricated and characterised to establish the material formulation landscape. Two response surface models (ReSMs) are subsequently developed, demonstrating that the proposed mixture design achieves a scalable range in both optical and mechanical properties, which is critical for VBTS design and fabrication. Given target material specifications, a desirability-function-based strategy is introduced to address the multi-objective tunability between transparency and hardness. The resulting theoretical optimum is further translated into a practical operating window, yielding manufacturable mixture compositions that can be directly realised via monolithic 3D printing and thereby completing the end-to-end inverse design pipeline.

To validate the effectiveness of the proposed i-Tac framework, comprehensive experiments were conducted, including material characterisation, theoretical analysis, fabricated sample evaluation, and functional VBTS performance assessment. The results demonstrate that, compared with conventional PolyJet materials, the proposed mixture elastomers substantially expand the scalable property space while narrowing the performance gap between 3D-printed elastomers and commercial silicones. Moreover, the combined use of response surface modelling and desirability optimisation enables accurate translation from target properties in the feasible property space to optimal material compositions. The framework was further employed to replicate the properties of representative commercial silicone elastomers by treating their characteristics as design targets, with the resulting samples exhibiting high property tunability. Finally, i-Tac demonstrates strong adaptability at the level of complete VBTS contact modules, successfully matching the characteristics of both custom-built and commercial sensors.

Overall, the proposed i-Tac inverse design framework provides a scalable pathway for fabricating VBTS elastomers with tunable optical and mechanical properties. This work lays an important foundation for advancing monolithic 3D printing of VBTS and for further exploration of high-performance tactile sensing systems.

\bibliographystyle{IEEEtran}

\bibliography{ref}

\newpage
\appendix

\subsection{Material Property Evaluation Details}

\subsubsection{Hardness}

For elastomers, softness is a key material attribute that distinguishes them from rigid solids. 
In tactile sensing applications, an elastomer with appropriate compliance can respond effectively to contact stimuli through measurable surface deformation. 
Hardness is commonly used as a quantitative metric (the inverse of softness), describing the resistance of a material to indentation or local deformation, and reflecting its resistance to permanent damage such as scratching, indentation, or cutting. In this work, an indentation-based Shore~00 hardness test is performed using a durometer to characterise the elastomers. During testing, the durometer was used to uniformly collect at least five data readings from the surface of each sample.

As summarised in Table~\ref{material property compare table}, the measured hardness values of the 21 materials span approximately 20-85~Shore~00 (very soft to soft). 
Among the 3D-printed materials, pure AC exhibits the highest stiffness, with an experimentally measured hardness of 84.8~Shore~00 (manufacturer specification: Shore~A~30). 
As the TM proportion in the AC/TM mixtures increases to 80\%, the hardness decreases proportionally to 43.2~Shore~00, indicating a substantial softening effect. 
A similar trend is observed in the GM/AC mixtures, although the softening occurs more rapidly, reaching 44.8~Shore~00 at 60\% GM. 
When AC is removed entirely, the TM/GM mixtures further reduce the hardness to approximately 20-30~Shore~00, corresponding to extremely compliant materials.

In the ternary AC/TM/GM mixtures, the hardness variation becomes more moderate due to the coupled material interactions. 
The multi-layer grid (SUP) sample exhibits a hardness of 68~Shore~00, representing a clear reduction from pure AC, comparable to AC6/TM4 and AC5/TM2/GM2.

For comparison, the five mold-cast silicones fall within the range of approximately 40-75~Shore~00, largely overlapping with the 3D-printed mixture designs. 
Ecoflex~00-31 and 00-45 are the softest (38.6/44.4~Shore~00), comparable to AC2/TM2/GM5 and GM6/AC4, respectively. 
DragonSkin~20 is significantly stiffer (approximately 70~Shore~00), similar to the SUP, while DragonSkin~30 and SORTA-Clear~40 further increase to around 75~Shore~00, positioned between AC8/TM2 and pure AC.

Overall, the results indicate that the proposed 3D-printed mixture designs can achieve softness levels comparable to mainstream PolyJet elastomers, such as AC, while offering flexible hardness tunability through compositional control comparative to mold-cast silicones.

\subsubsection{Poisson's Ratio}

Poisson's ratio $v$ is a fundamental mechanical parameter describing the relationship between transverse strain $\varepsilon_{\text{trans}}$ and axial strain $\varepsilon_{\text{axial}}$ under uniaxial loading. 
In tactile sensing applications, near-incompressibility ($v=0.5$) implies that geometric deformation under external loading occurs with limited volume variation.

\begin{equation}
v = -\frac{d\varepsilon_{\text{trans}}}{d\varepsilon_{\text{axial}}}
\end{equation}

In this study, Poisson's ratio is estimated using Digital Image Correlation (DIC). 
During uniaxial tensile testing, surface markers applied to ASTM~D412 dog-bone specimens are tracked to obtain full-field displacement data. 
Axial and transverse strains are then extracted from the measured marker displacements, from which $v$ is calculated. Although DIC provides a non-contact and spatially resolved strain measurement approach, its finite spatial resolution and sensitivity to surface texture, illumination conditions, and speckle quality may introduce uncertainty. Consequently, the experimentally obtained Poisson's ratios may deviate from their theoretical values.

As summarised in Table~\ref{material property compare table}, the measured Poisson's ratios of the 21 samples range from 0.35 to 0.45. 
For the 3D-printed elastomers, most values fall within 0.40-0.45, approaching the commonly assumed incompressible limit and indicating relatively low volumetric compressibility. By comparison, the mold-cast silicones exhibit measured values between 0.35 and 0.40, whereas their nominal Poisson's ratios are typically reported in the range of 0.47-0.50. 
This systematic underestimation is primarily attributed to DIC-related measurement uncertainty and practical material variability. 
Importantly, both the 3D-printed mixture designs and the mold-cast silicones are evaluated using the same DIC setup and processing pipeline; therefore, the relative comparison between the two material groups remains valid. 
Under this consistent measurement framework, the mixture-designed elastomers demonstrate measured incompressibility that is comparable to, and in most cases slightly higher than, that of the commercial silicones.

\subsubsection{Uniaxial Tensile Experiment}

Uniaxial tensile testing is widely used to characterise the mechanical behaviour of elastomers under unidirectional loading. 
Its primary objective is to obtain stress-strain curves for analysing tensile properties such as elastic modulus and tensile hysteresis. 
ASTM~D412 is a well-established standard for evaluating the tensile properties of elastomers, specifying specimen geometry, testing conditions, and reporting procedures. For VBTS tactile sensing, shear contact dynamic is closely related to the material tensile property. The tensile test results are presented in Fig.~\ref{material property fig}(A).

\begin{itemize}

\item \textbf{Monotonic Tensile Loading:} 
Fig.~\ref{material property fig}(A.a) shows the stress $\sigma$-strain $\varepsilon$ responses of the 21 elastomers under monotonic loading from 0 to 100\% strain. 
Most materials exhibit approximately linear behaviour within this range; however, AC displays noticeable nonlinearity. 
Specifically, the stress of AC increases rapidly up to approximately 10\% strain, after which the growth rate gradually decreases and becomes nearly linear beyond 20\% strain. 
AC also consistently exhibits higher stress levels than the other materials throughout the loading process. The stress-strain curves of the remaining mixture designed elastomers generally fall within the envelope defined by the mold-cast silicones.

To quantify tensile stiffness, the engineering stress at 100\% strain is used to approximate the elastic modulus $E$ (Young's modulus), which characterises the material resistance to tensile deformation. 
Due to minor measurement fluctuations near 100\% strain, the peak stress immediately prior to the fluctuation is adopted as the representative value of $E$.

\begin{equation}
E = \frac{\sigma_{100\%}}{\varepsilon_{100\%}} = \sigma_{100\%}
\end{equation}

As summarised in Table~\ref{material property compare table}, the elastic modulus of the 21 materials spans approximately 48-450~kPa, representing nearly an order-of-magnitude variation. 
Consistent with Fig.~\ref{material property fig}(A.a), the traditional PolyJet material AC, exhibits the highest modulus (449.87~kPa), approximately 70\% higher than the SUP (270~kPa).

In comparison, for the 3D-printed mixture designs, both AC/TM and GM/AC elastomers show a gradual reduction in modulus from approximately 230~kPa to 130~kPa as the AC fraction decreases from 80\%. While TM/GM mixtures occupy a substantially lower range (approximately 80-90~kPa), corresponding to roughly one-fifth of the stiffness of pure AC.
Furthermore, the mold-cast silicones are distributed across three characteristic levels: approximately 330~kPa (SORTA-Clear 40), 235~kPa (DragonSkin~20), and about 50~kPa (Ecoflex). Notably, the softest 3D-printed materials has closely reached commercial silicone through the lowest modulus observed from mixture designs.

\item \textbf{Cyclic Tensile Loading}: 
Cyclic tensile tests are conducted to evaluate tensile hysteresis, which reflects the internal energy dissipation between loading and unloading within a stretch cycle. 
The hysteresis is quantified by the normalised area difference between the loading and unloading stres-strain curves. A larger ratio indicates greater energy dissipation and more pronounced viscoelastic effects.

\begin{equation}
H = \frac{A_{\text{load}} - A_{\text{unload}}}{A_{\text{load}}}
\end{equation}

The quantitative results in Table~\ref{material property compare table} indicates that the tensile hysteresis of AC and the SUP reach 0.64 and 0.33, respectively. For the mixture-designed elastomers, hysteresis generally decreases with reduced AC content. Most mixtures containing less than 40\% AC achieve hysteresis values below 0.1, comparable to those of the mold-cast silicones.

\item \textbf{Combined Tensile Performance:}
The combined relationship of elastic modulus and tensile hysteresis is illustrated in Fig.~\ref{material property fig}(A.b). AC occupies the upper-right region of the plot, characterised by both high stiffness and high energy dissipation, indicating strong load resistance but relatively slow dynamic response. 
The SUP follows a similar trend but with both metrics reduced to approximately half of those of pure AC.

For the remaining 3D-printed mixture designs, the AC/TM mixtures show an approximately proportional relationship between AC content and both modulus and hysteresis, forming a near-linear distribution. 
In GM/AC mixtures, a similar scaling trend is evident primarily in the elastic modulus, whereas the hysteresis values cluster within the range of 0.2-0.3. 
TM/GM mixtures exhibit substantially reduced modulus ($\lesssim$100~kPa) and hysteresis ($\lesssim$0.1), indicating rapid response but without a clear linear trend. The ternary AC/TM/GM mixtures demonstrate tensile behaviour comparable to representative AC/TM compositions (e.g., AC2/TM8 and AC4/TM6).

By comparison, the mold-cast silicones show a more horizontally distributed pattern, suggesting that they can provide a wide stiffness range while maintaining relatively stable hysteresis. 
In contrast, the tensile performance of the 3D-printed elastomers remains partly constrained by the intrinsic viscoelastic characteristics of AC, such that achieving higher stiffness tends to be accompanied by increased hysteresis. However, from a comparative perspective, mixture design demonstrates a marked improvement in tensile material properties relative to conventional PolyJet materials such as AC and SUP.

\end{itemize}

\subsubsection{Uniaxial Compressive Experiment}

In practical tactile sensing scenarios, the dominant loading mode is compression, as normal force is required to establish effective contact. 
Consequently, the elastomer layer of the tactile sensor primarily undergoes compressive deformation during operation. 
Understanding the compressive behaviour of candidate materials is therefore essential for sensor design and performance optimisation. 
ASTM~D575 provides a standardised protocol for evaluating the compressive properties of rubber and elastomeric materials under uniaxial loading, and is adopted here to characterise both 3D-printed and mold-cast elastomers. 
The experimental results are presented in Fig.~\ref{material property fig}(B).

\begin{itemize}

\item \textbf{Monotonic Compression:}
Fig.~\ref{material property fig}(B.a) shows the stress-strain responses under monotonic compression up to 20\% strain. 
Unlike the tensile results, DragonSkin~30 defines the upper boundary of the response envelope rather than AC. 
The two materials exhibit similar behaviour below approximately 5\% strain; however, with increasing compression depth, DragonSkin~30 demonstrates a noticeably steeper stress growth than AC. 
SORTA-Clear~40 and DragonSkin~20 follow trends similar to DragonSkin~30 and AC, respectively, and both exceed the stress of AC beyond approximately 15\% strain.

The remaining materials exhibit more gradual stress increases, with TM/GM mixtures forming the lower boundary of the response range. 
To quantify compressive stiffness, the engineering stress at 10\% strain is defined as the compressive modulus, as summarised in Table~\ref{material property compare table}. 
The mold-cast silicones show generally higher compressive modulus values, forming two characteristic bands (approximately 35-50~kPa and 170-230~kPa). 
In contrast, the 3D-printed mixture designs span a broader but overall lower range (approximately 7-150~kPa), indicating reduced resistance to compressive loading. This demonstrates that in VBTS tactile sensing, mixture design elastomers can provide force sensitivity comparable to that of mold-cast silicone.

\item \textbf{Cyclic Compression:} 
Cyclic compression tests are performed to evaluate compressive hysteresis, which characterises the energy dissipation during loading-unloading cycles. 
As with the tensile case, hysteresis is quantified by the normalised area difference between the loading and unloading curves.

The quantitative results in Table~\ref{material property compare table} shows that AC exhibits pronounced compressive hysteresis despite its moderate compressive modulus, indicating substantial viscoelastic energy dissipation and relatively slow recovery dynamics. Similarly, AC8/TM2 reach hysteresis values of approximately 0.5, followed by AC6/TM4, TM8/GM2, and GM4/AC6 at around 0.4, and AC4/TM6, GM6/AC4, and GM2/AC8 at approximately 0.37.

In comparison, most of the remaining 3D-printed mixtures and the mold-cast silicones fall within the range of 0.15-0.30, indicating faster dynamic recovery than typical PolyJet material, AC.

\item \textbf{Combined Compressive Performance:}
The combined relationship of compressive modulus and hysteresis is illustrated in Fig.~\ref{material property fig}(B.b). 
Overall, clear differences are observed between the 3D-printed and mold-cast elastomers. 
The 3D-printed materials generally occupy a lower modulus region (below approximately 150~kPa) but exhibit a wider hysteresis spread (approximately 0.1-0.5). 
Within the mixture designs, the AC and TM fractions in AC/TM and TM/GM systems show observable correlations with compressive behaviour, whereas the influence of GM is less pronounced.

By contrast, the mold-cast elastomers extend to higher compressive modulus values (up to approximately 230~kPa) while maintaining relatively stable hysteresis (0.1-0.25). 
This suggests that commercial silicones may be more suitable for applications requiring high compressive stiffness (e.g., DragonSkin and SORTA-Clear), but also with lower normal force sensitivity. For applications favouring lower compressive modulus, the 3D-printed mixture designs can achieve dynamic responses comparable to Ecoflex while offering flexible property scalability (e.g., AC/TM/GM, TM4/GM6, and AC2/TM8).

\end{itemize}

\end{document}